\documentclass[a4paper,onecolumn,allowtoday, accepted=2026-04-03]{quantumarticle}
\pdfoutput=1
\usepackage{orcidlink} 
\usepackage[utf8]{inputenc}
\usepackage[toc,page]{appendix}
\usepackage{amsfonts}
\usepackage{color}
\usepackage{amsmath}
\usepackage{graphicx}
\usepackage{enumitem}
\usepackage{subcaption}
\usepackage{matlab-prettifier}
\usepackage{algorithm}
\usepackage{algpseudocode}
\usepackage[many]{tcolorbox}
\usepackage[framemethod=default]{mdframed} 
\usepackage{showexpl}
\definecolor{DarkRed}{rgb}{.7,.1,.1}
\definecolor{DarkBlue}{rgb}{.11,.23,.60}
\definecolor{LightBlue}{rgb}{.3,.3,.8}
\mdfdefinestyle{exampledefault}{%
rightline=true,innerleftmargin=10,innerrightmargin=10,
frametitlerule=true,
backgroundcolor=black!10!white,
frametitlerulecolor=blue,
frametitlefont={\color{white}},
frametitlebackgroundcolor=DarkBlue,
frametitlerulewidth=1pt}

\usepackage{amssymb}
\usepackage{hyperref}
\usepackage{color}
\usepackage{caption}
\usepackage{mathrsfs}
\usepackage{isomath}
\usepackage{amsthm}
\usepackage{epstopdf}
\usepackage{txfonts}
\usepackage{dsfont}
\usepackage{physics} 
\usepackage{tikz} 
\allowdisplaybreaks[4]

\usepackage{bbold}
\usepackage{mathtools}

\newcommand{\aver}[1]{\langle #1 \rangle}

\renewcommand{\vec}[1]{\boldsymbol{#1}}

\newcommand{\id}{\mathds{1}}
\renewcommand{\ket}[1]{| #1 \rangle}
\renewcommand{\bra}[1]{\langle #1 |}

\newcommand{\de}{\mathrm{d}}

\newcommand{\kb}[2]{| #1 \rangle\langle #2 |}

\newcommand{\updm}{\ketbra{\tfrac{1}{2}}}
\newcommand{\downdm}{\ketbra{-\tfrac{1}{2}}}

\newcommand{\Sep}{{\rm SEP}}

\newcommand{\BSA}{\mathcal E_{BSA}}

\newcommand{\be}{\begin{equation}}
\newcommand{\ee}{\end{equation}}
\newcommand{\eea}{\end{eqnarray}}
\newcommand{\bea}{\begin{eqnarray}}

\renewcommand{\va}[1]{\ensuremath{(\Delta#1)^2}}

\newcommand{\EW}{\ensuremath{\mathcal{W}}}
\newcommand{\ignore}[1]{}

\usepackage{cleveref}
\crefname{equation}{Eq.}{Eqs.}
\crefname{observation}{Obs.}{Obs.}
\crefname{corollary}{Corollary}{Corollaries}
\crefname{lemma}{Lemma}{Lemmata}
\crefname{proof}{Proof}{Proofs}
\creflabelformat{proof}{#2proof#3}
\crefname{remark}{Remark}{Remarks}
\crefname{prop}{Proposition}{Propositions}
\begin{document}

\title{Estimating the best separable approximation of non-pure spin-squeezed states}

\author{Julia Math\'e \,\orcidlink{0009-0002-7403-7044}}
\email{julia.mathe@tuwien.ac.at}
\affiliation{Vienna Center for Quantum Science and Technology, Atominstitut, TU Wien,  1020 Vienna, Austria}

\author{Ayaka Usui \,\orcidlink{0000-0002-2326-3917}}
\email{ayaka.usui@uab.cat}
\affiliation{Departament de F\'{i}sica, Universitat Aut\`{o}noma de Barcelona, 08193 Bellaterra, Spain}

\author{Otfried G\"uhne \,\orcidlink{0000-0002-6033-0867}}
\email{otfried.guehne@uni-siegen.de}
\affiliation{Naturwissenschaftlich-Technische Fakult{\"a}t, Universit{\"a}t Siegen, Walter-Flex-Stra{\ss}e 3, D-57068 Siegen, Germany}

\author{Giuseppe Vitagliano \,\orcidlink{0000-0002-5563-3222}}
\email{giuseppe.vitagliano@tuwien.ac.at}
\affiliation{Vienna Center for Quantum Science and Technology, Atominstitut, TU Wien,  1020 Vienna, Austria}

\date{\today}

\begin{abstract}
We discuss the estimation of the distance of a given mixed many-body quantum state to the set of fully separable states, applied to the concrete scenario of collective spin states. Concretely, we discuss \emph{lower bounds} to distances from the set of fully separable states based on entanglement criteria and \emph{upper bounds} to those distances using an iterative algorithm to find the optimal separable state closest to the target. Focusing on 
collective states of \texorpdfstring{\(N\)}{N} spin-\texorpdfstring{\(1/2\)}{1/2} particles, we consider spin-squeezing inequalities (SSIs), which provide a complete set of nonlinear entanglement criteria based on collective spin variances. First, we find a lower bound to distance-based entanglement monotones, specifically the so-called best separable approximation (BSA) from the complete set of SSIs, thereby bypassing entirely a numerical optimization over a (potentially very large) set of linear entanglement witnesses. Then, we improve current algorithms to iteratively find the closest separable state to a given target state, exploiting the symmetry of the system. These results allow us to study entanglement quantitatively on thermal states of spin systems on fully-connected graphs at \emph{nonzero} temperature, as well as potentially similar states arising in out-of-equilibrium situations. We thus apply our methods to investigate entanglement across different phases of a fully-connected XXZ model. We observe that our lower bound becomes often tight for zero temperature as well as for the temperature at which entanglement disappears, both of which are thus precisely captured by the SSIs. We further observe, among other things, that entanglement can arise at nonzero temperature even in the ordered phase, where the ground state is separable, revealing the potential usefulness of entanglement quantification also beyond the ground state paradigm.
\end{abstract}

\maketitle

\section{Introduction}

Entanglement distinguishes genuine quantum correlations from classical ones and serves as an important resource in quantum information tasks~\cite{HorodeckiEntanglementReview2009}, as well as a tool for understanding complex many-body quantum systems~\cite{amico08}. Notably, when analyzing complex phases of matter, the entanglement structure of the zero-temperature phase diagram shows interesting features, especially across quantum phase transitions (QPTs)~\cite{amico08,Laflorencie16,sachdev,girvin_yang_2019}. In fact, the scaling of ground-state entanglement has immediate implications for numerical simulation methods~\cite{SchollwockRev,Schollwoeck2011,Orus2019}. 
However, entanglement is challenging to characterize in practical situations, when the quantum state is noisy and only very partially known~\cite{GuehneToth2009, FriisVitaglianoMalikHuberReview19}. Generally, even for fully known density matrices, such as simple nonzero temperature equilibrium states, just deciding whether a state is entangled or not quickly becomes computationally unfeasible with increasing particle numbers.

\begin{figure}[ht]
 \begin{center}
    \includegraphics[width=1\textwidth]{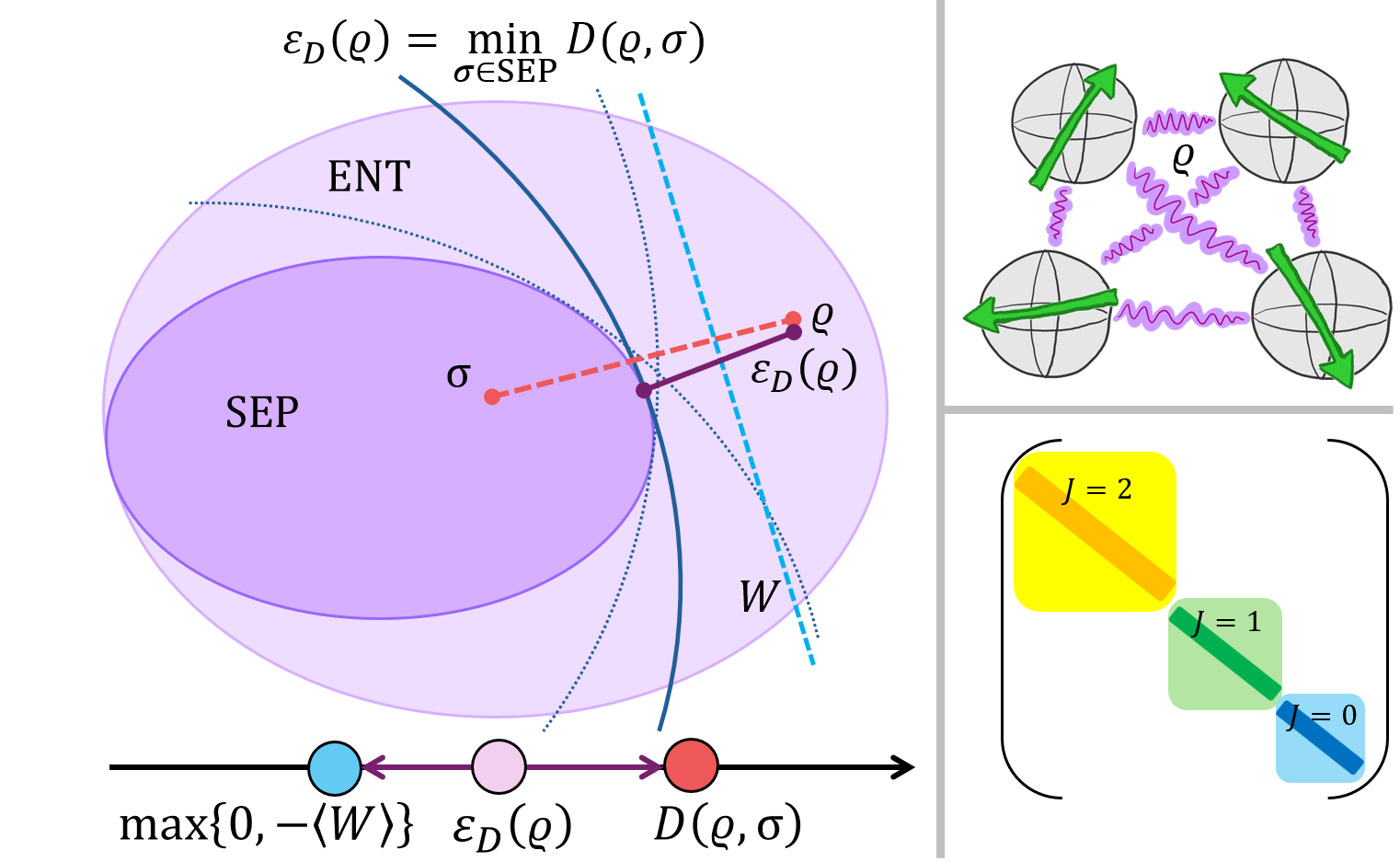}
    \end{center}
    \caption{We estimate an entanglement monotone $\mathcal{E}(\varrho)$, given by the minimal distance to the set of separable states, for an $N$-qubit state (top right), by computing lower and upper bounds (left). Lower bounds are computed from the negative expectation value of an entanglement witness $W$ that is illustrated as blue lines not crossing the set of separable states (shaded ellipse). Upper bounds can be computed from the distance to any given separable state (red/purple lines), minimized over as many separable states as possible. Optimal bounds are illustrated as straight or dotted lines, non-optimal bounds as dashed lines. In particular, the dotted blue lines illustrate the optimization over a set of optimal witnesses, which is exactly what happens in the case of the SSIs. For states featuring symmetries, the problem can be formulated in a symmetric subspace (example matrix on the bottom right: for $N = 4$, blocks of different color are a result of permutational invariance, while diagonal matrices have an additional rotation symmetry).}
    \label{fig: overview}
\end{figure}

In many-particle states, entanglement has been characterized quantitatively mostly in pure ground states, typically between two particles at growing distance, or across bipartitions of increasing size~\cite{amico08,Laflorencie16}. 
In pure states, bipartite entanglement (across a bipartition or between two particles) can be quantified via entropies of one of the marginals.  
Those are instances of {\it entanglement monotones}~\cite{Vedral_1998, PlenioVirmani07,HorodeckiEntanglementReview2009}, that are nonnegative real functions $\mathcal{E}(\varrho)$ that vanish for all separable states and do not increase under local operations and classical communication (LOCC). 
Beyond pure states (especially ground states), the exact values of entanglement monotones are known only in very special cases, e.g., two-qubits~\cite{wootters2qubitsEoF}, highly symmetric states of two particles~\cite{wootters2qubitsEoF,VollbrechtWerner2001}, or special entanglement measures of three qubits~\cite{UhlmannPRA2000,LohmayerPRL2006,Eltschka_2008,OsterlohSiewertUhlmannPRA08,SiewertEltschkaPRL12}. One exception that works in the many-body regime is represented by the entanglement negativity, which can be calculated in bosonic Gaussian states~\cite{audenaert2002,amico08,eisert2010,calabrese2013entanglement,Eisler_2014,calabrese2015finite}, and estimated for fermionic ones~\cite{eisler2015partial,EislerZimborasPRB2016,EisertEislerZimboras_2018}. However, the latter measure is not faithful, in the sense that entangled states with a positive partial transpose (PPT) have zero negativity. Such PPT entangled states are quite ubiquitous~\cite{HorodeckiEntanglementReview2009,GuehneToth2009,marconi2025symmetricquantumstatesreview}, especially in the many-body scenario, in which they also
arise as thermal states~\cite{Ferraro_2008,Cavalcanti_2008,tothPRA09,Vitagliano2025sudsqueezingmany}. An alternative approach that is applicable to mixed states, consists in considering {\it entanglement witnesses}~\cite{GuehneToth2009, FriisVitaglianoMalikHuberReview19}, which allow to tell whether a state is entangled or not, and potentially could provide bounds to entanglement monotones~\cite{brandao05,Audenaert06,EisertBrandaoAudenaert07,GuhneReimpellWernerPRL07,GuhneReimpellWerner08,FadelVitagliano_2021}. 
This idea has in fact been explored in a variety of systems~\cite{tothpra05,anderswinter2008,amico08,marty14,marty16,Pappalardi_2017,Lerose_2020,IgloiTothPRR2023,Mazza_2025}. Entanglement witnesses have been also associated to thermodynamically relevant quantities such
as energy~\cite{DowlingDohertyBartlett2004,tothpra05,Anders_2006}, heat capacity \cite{WiesniaketalPRB2008}, magnetic susceptibilities ~\cite{Wiesniak05,brukner06} or generic response functions~\cite{krammer09,marty14,HaukeHeylTagliacozzoZoller16,Mazza_2025}, and can also allow to distinguish genuine multipartite entanglement~\cite{GuhneTothPRA2006EnergyandMultipartite}.

However, so far there is still a lack of a {\it systematic} extension of the quantitative study of entanglement in many-body systems
from the pure-state paradigm to the physically ubiquitous case of mixed states. We address this problem in a setting that is both of high physical relevance and sufficiently
structured to allow an \emph{ab-initio} investigation, which is given by collective spin models. Those systems are of paradigmatic interest in quantum many-body physics, as they capture typical interaction mechanisms in ensembles of spins or effective two-level atoms, and exhibit nontrivial equilibrium and dynamical behavior. Importantly, these systems possess a high degree of symmetry (e.g.\ permutation and rotation symmetry) that can be
exploited to reduce complexity and to make quantitative entanglement analysis possible even for large $N$.

Because of this, long-range spin systems are frequently treated using mean-field or
semi-classical approaches based on spin-coherent states, and/or by restricting attention to pure states or to a
single total-spin-$j$ sector~\cite{Ribeiro_2007,Ma2011Quantum,DEFENU20241}. While such approaches provide valuable intuition, 
as well as precise quantitative predictions in the semi-classical regime, they can miss qualitatively
new regimes where quantum correlations are strong or where multiple $j$-subspaces contribute~\cite{Vidal2004Entanglement_first,mann2025squeezingclassicalantiferromagnetsquantum}, e.g., due to mixing. 
As a result, much remains to be explored already within these highly
symmetric models, especially \emph{outside} the domain of validity of semi-classical spin physics and simple
mean-field descriptions.

As mentioned earlier, a key concept that helps estimate entanglement is given by entanglement witnesses. These are observables $W\in \EW$ such that $\tr(W\sigma)\geq 0$ for all separable states $\sigma$ and there exist entangled states $\varrho$ for which $\tr(W\varrho) < 0$.
Important examples for collective spin states, and in general for many-particle systems are closely connected to uncertainty relations and metrology, which in this context translates into spin squeezing~\cite{KitagawaUeda1993,SoerensenNAT2001,SoerensenMoelmer2001,tothPRL07,tothPRA09,Ma2011Quantum,vitagliano11,vitagliano14,vitagliano16,MartyVitagliano}.
In this setting, the key advantage is that such witnesses can be expressed in terms of first and second moments of collective spin operators, which makes them readily calculable in many-spin models~\cite{Ma2011Quantum}, and also measurable experimentally. In fact, spin-squeezing witnesses are also nowadays routinely implemented experimentally, especially in atomic gases~\cite{PezzeRMP2016,FriisVitaglianoMalikHuberReview19}. In particular, a complete set of SSIs has been derived~\cite{tothPRL07, tothPRA09} and later extended to higher spin ensembles~\cite{vitagliano11,vitagliano14,Vitagliano2025sudsqueezingmany}.
This set can be represented as a polytope in the space of first and second collective spin moments, which is fully filled by separable states in the limit of large $N$.
States outside this polytope must be entangled, whereas points on its boundary are attained by separable states, which can be regarded as generalized spin-coherent states. Because of that, this complete set of generalized SSIs can be considered a faithful characterization of entangled collective spin states, at least those characterized by fluctuations over semi-classical (i.e., spin-coherent) states. Thus, the question arises whether this set provides witnesses that are optimal in a {\it quantitative} sense, i.e., providing tight lower bounds to entanglement monotones, which is what we explore in this work. 

Guided by these considerations, the key focus of this work is to develop and apply methods that enable
quantitative entanglement characterization in mixed collective-spin states in a computationally feasible
way. As quantifiers of entanglement, we consider distances from the set of fully-separable states,
which provide a faithful characterization of entanglement and are agnostic to the different forms of entanglement (e.g., bipartite versus genuine multipartite).
Moreover, these measures can be more directly bounded from entanglement witnesses~\cite{brandao05}, also those related to spin squeezing~\cite{FadelVitagliano_2021}. 
In particular, we then ask how to obtain both: (i) a (possibly tight) lower bound to the distance to the set of fully-separable states (exploring different possible metrics) via optimal witnesses, in particular the complete set of generalized SSIs of Refs.~\cite{tothPRL07, tothPRA09}, and (ii) an (iterative) estimate to the closest fully-separable state, 
in particular an exact separable decomposition in the case of fully separable states. The latter goal is very important to provide a faithful characterization of entangled versus separable states.
Furthermore, we aim at reaching our goals in a way that scales favorably with system size by exploiting collective structure and symmetry.
This provides a route to move beyond ground-state-only analyses in equilibrium physics and beyond pure-state
treatments of out-of-equilibrium processes, and hence towards a more systematic mixed-state entanglement theory for long-range spin systems.

Concretely, our main contributions are as follows:

\begin{enumerate}[label=\Alph*]
    \item We derive a simple closed-form optimal spin-squeezing parameter from the complete set of generalized SSIs, from which we bound the distance to the set of fully-separable states.
    \item We improve iterative algorithms that find closer and closer separable ansatz states, thereby providing upper bounds from such distances. 
    \item We apply these methods to thermal states on the fully-connected XXZ model and reveal how a distance-based entanglement monotone, the best separable approximation (BSA), behaves across different phases and phase transitions, approaching them from nonzero temperatures. With this numerical investigation we observe that: 
    \begin{enumerate}
        \item It is possible to quantify generalized spin squeezing and thereby get a lower bound to the BSA for a large number of particles, or even analytically for large $N$ in some cases.
        \item The generalized spin-squeezing parameter does in fact often faithfully capture entanglement in thermal equilibrium states of these quadratic models.
        \item The separable decomposition can be found, after exploiting the symmetry of the state, for a significantly higher number of qubits ($\sim 10$) with respect to current limits on generic states ($3$ or $4$).
        \item Entanglement arises at nonzero temperature when the ground state is separable (i.e., in an ordered phase) and the parameters are not far from the point where a QPT would arise in the thermodynamic limit, thus potentially revealing
        connections between entanglement and quantum criticality beyond ground states.
    \end{enumerate}
\end{enumerate}

Note that the fully-connected XXZ model fits well in our framework and has a relatively rich phase diagram, containing QPTs both in the ferromagnetic (FM) and anti-ferromagnetic (AFM) cases. Note also that the phase diagram of the model is different when all spin-$j$ subspaces of the $N$-spin-$1/2$-particle system (with $0\leq j \leq N/2$) are considered~\cite{Vidal2004Entanglement_first}\footnote{Throughout this work, QPTs refer to the thermodynamic limit. While $N \rightarrow \infty$ is analytically accessible in some cases, in all our numerical analysis we can only consider finite-size systems.}.

The paper is structured as follows. In \cref{sec:methods} we introduce the relevant methods: In \cref{sec:ent_dist_monotones} we elaborate on entanglement quantification in many-body systems, specifically focusing on permutationally invariant systems, and the so-called best separable approximation (BSA) as an entanglement quantifier. In \cref{sec:spin_squeezing_section}, we give an overview of spin-squeezing and explain how this notion yields powerful collective-spin witnesses and why they make potentially good candidates for lower bounds to the BSA and other monotones. In \cref{sec:ansatzintro}, we explain how to find upper bounds using an iterative ansatz to find the closest separable state, and how symmetries may be useful in improving this ansatz. In \cref{sec:perm_inv_XXZ}, we introduce the permutationally invariant XXZ model at thermal equilibrium as the paradigmatic model of interest. Afterwards, in \cref{sec:results} we apply our methods to this concrete setting and discuss our results in this context. First, in \cref{subsec:SSIs_BSA}, we define a single compact spin squeezing parameter that incorporates the optimal SSI witness for a given collective spin state and show how to lower bound the BSA from this parameter. In \cref{sec:upper_bound_symm} we explicitly show how to make use of the symmetry to improve the search of an ansatz state that provides a good upper bound. Finally, in \cref{sec:numerical_results} we present numerical simulations on thermal states of the fully connected XXZ model, showing that our method can be effectively implemented and provides a good estimate of entanglement, especially hinting at the fact 
that SSIs do detect most, if not all entangled states in that context. In some special cases, we are also able to obtain {\it analytically} the asymptotic scaling of the spin-squeezing parameter, and thereby a potentially tight lower bound to the BSA in the limit $N\rightarrow \infty$, which we describe in \cref{sec:analytic}. Finally, in \cref{sec:conclusions} we draw our conclusions and discuss some outlook on the applications of our methods in out-of-equilibrium scenarios and elaborate on possible future works. 

\section{Methods}\label{sec:methods}

\subsection{Definition of entanglement and its quantification}
\label{sec:ent_dist_monotones}

Let us consider a system of $N$-qubits. A state is \emph{fully separable} if it can be decomposed as
\begin{equation}
\label{eq:sep_def}
\varrho_{\rm sep}=\sum_k p_k \left(\varrho_1\otimes\cdots\otimes\varrho_N\right)_k ,
\end{equation}
where each $\varrho_j$ is a single-particle density matrix and $\{p_k\}_k$ is a probability distribution.
The set of fully separable states is convex and closed, and we denote it by $\Sep$.
A density matrix $\varrho$ is \emph{permutationally invariant} (PI) if
\begin{equation}
U_\pi \varrho U_\pi^\dagger=\varrho \qquad \forall\,\pi\in\mathfrak{S}_N ,
\end{equation}
where $U_\pi$ is the unitary representation of the permutation $\pi$ and $\mathfrak{S}_N$ denotes the set of permutations of $N$ parties.
A convenient way to enforce PI is via the \emph{PI twirling} map~\cite{VollbrechtWerner2001,EggelingWerner2001}
\begin{equation}
\label{eq:PI_twirl_map}
\mathcal{T}_{\rm PI}(\cdot):=\frac{1}{N!}\sum_{\pi\in\mathfrak{S}_N}U_\pi(\cdot)U_\pi^\dagger .
\end{equation}
If $\varrho_{\rm sep}$ is separable, then $\mathcal{T}_{\rm PI}(\varrho_{\rm sep})$ is both separable and PI, and we denote the (convex) set of such states by $\Sep_{\rm PI}:=\Sep\cap\{\text{PI states}\}$.

It is useful to characterize the \emph{boundary} of $\Sep_{\rm PI}$ through PI-twirled pure product states.
Let
\begin{equation}
\ketbra{\Psi_{\rm prod}}=\ketbra{\psi_1}\otimes\cdots\otimes\ketbra{\psi_N}
\end{equation}
be a pure product state. Its PI twirl,
\begin{equation}
\label{eq:prod_PI_boundary}
\varrho_{\rm prod,PI}:=\mathcal{T}_{\rm PI}\!\left(\ketbra{\Psi_{\rm prod}}\right)
=\frac{1}{N!}\sum_{\pi\in\mathfrak{S}_N}U_\pi\left(\ketbra{\psi_1}\otimes\cdots\otimes\ketbra{\psi_N}\right)U_\pi^\dagger ,
\end{equation}
typically mixed, is at the boundary of $\Sep_{\rm PI}$~\cite{VollbrechtWerner2001,EggelingWerner2001}. Note also that $\ket{\Psi'_{\rm prod}}=U_\pi\ket{\Psi_{\rm prod}}$ yields the same $\varrho_{\rm prod,PI}$.
In the special case of \emph{permutationally symmetric} states (supported on the totally symmetric subspace), one can further restrict to mixtures of symmetric pure product states with identical single-particle factors (cf. \cref{app:separablePIstates}). A similar twirling applies to other local symmetries, for example under rotations~\cite{EggelingWerner2001}, see \cref{app:separablePIstates} for a more detailed discussion. 

A state is \emph{entangled} if it is not separable, i.e.\ $\varrho\notin\Sep$.
A standard tool to detect entanglement is an \emph{entanglement witness}: a Hermitian operator $W$ such that
\begin{equation}
\label{eq:witness_def}
\tr(W\sigma)\ge 0\quad \forall\,\sigma\in\Sep,
\qquad\text{and}\qquad
\exists\,\varrho\notin\Sep:\ \tr(W\varrho)<0 .
\end{equation}
Geometrically, witnesses define supporting hyperplanes of the convex set $\Sep$ (see, e.g., the review in~\cite{HorodeckiEntanglementReview2009,GuehneToth2009,FriisVitaglianoMalikHuberReview19} and \cref{fig: overview}(left) for an illustration).
Importantly for our setting, if the state $\varrho$ possesses symmetries, witnesses detecting it can be restricted to those sharing the same symmetries, without loss of optimality (see, e.g., ~\cite{brandao05,Audenaert06,EisertBrandaoAudenaert07,GuhneReimpellWernerPRL07,GuhneReimpellWerner08}).
In particular, for PI, or rotation invariant states, we may restrict to PI or rotation invariant witnesses.

Beyond detection, one can quantify entanglement via \emph{entanglement monotones} $\mathcal{E}(\varrho)$: nonnegative functionals such that (i) $\mathcal{E}(\sigma)=0$ for all $\sigma\in\Sep$ and (ii) $\mathcal{E}$ does not increase under LOCC; in particular $\mathcal{E}$ is invariant under local unitaries (see, e.g.,~\cite{PlenioVirmani07,HorodeckiEntanglementReview2009}).
A broad class of monotones can be defined as an optimization over (subsets of) witnesses~\cite{brandao05}:
\begin{equation}
\label{eq:defWE}
\mathcal{E}_{\mathcal{M}}(\varrho)
=\max\Bigl\{0,\ -\min_{W\in\mathcal{M}}\tr(W\varrho)\Bigr\},
\end{equation}
where $\mathcal{M}$ is a chosen subset of entanglement witnesses.
This form makes the convex duality between separable states and entanglement witnesses transparent and, for many relevant choices of $\mathcal{M}$, makes it easy to find lower bounds~\cite{brandao05,Audenaert06,EisertBrandaoAudenaert07,GuhneReimpellWernerPRL07,GuhneReimpellWerner08}, as we will clarify below.

A complementary viewpoint is to quantify entanglement as a \emph{distance}\footnote{Note that one does not require $D(\varrho, \sigma)$ to be symmetric, thus also pseudo-distances can be considered.} to the separable set~\cite{Vedral_1997}:
\be\label{eq:distmeasgen}
\mathcal E_D(\varrho) =  \min_{\sigma \in {\rm SEP}} D(\varrho , \sigma) .
\ee
Common examples of such entanglement monotones are the relative entropy of entanglement~\cite{Vedral_1998,VollbrechtWerner2001}
or distance measures based on, e.g., the Uhlmann fidelity, such as the geometric measure of entanglement~\cite{streltsov2010linking}\footnote{Note that although this measure was introduced for pure states, it can also be extended to mixed states via a convex-roof construction, and not directly as a minimization over separable states.}.

From a many-body perspective, they admit a particularly transparent interpretation in terms of the mean-field approximation: they quantify \emph{how far} the quantum many-body state is from the closest fully separable state, i.e., from the optimal semi-classical description within a mixed product-state ansatz.
In other words, they measure the minimal ``correction'', in a chosen metric, needed to go beyond a mean-field state. 

As illustrated in \cref{fig: overview}, our strategy is to estimate $\mathcal{E}(\varrho)$ by combining (i) lower bounds obtained from entanglement witnesses and (ii) upper bounds obtained by constructing explicit separable approximations to $\varrho$.
While the former amounts to finding better and better entanglement witnesses (i.e., optimizing over all of them eventually), the latter amounts to searching for separable states that are progressively closer to $\varrho$.
It is known that in principle entanglement monotones can be lower bounded from {\it any} given entanglement witness $W$ by making use of convex duality~\cite{Audenaert06,EisertBrandaoAudenaert07,GuhneReimpellWernerPRL07,GuhneReimpellWerner08}. 
Abstractly speaking, such a bound, which we denote as $\mathcal E^{\rm low}(\varrho , W)$ (which is typically hard to find in the first place) could even be optimized over all (or a subset of) entanglement witnesses so that one has
\be\label{eq:elowboundgen}
\mathcal E(\varrho) \geq \max_{W \in \EW} \mathcal E^{\rm low}(\varrho , W) .
\ee
Once again, in general finding such a bound, not to mention optimizing over a large set of witnesses, is very challenging, especially in a many-particle scenario. 

On the other hand, for the entanglement monotones in \cref{eq:defWE}, that are directly defined via optimization over a subset of witnesses finding a lower bound amounts to calculating
the expectation value of {\it any} witness in the given set, $\mathcal M$. In other words, for measures defined as in \cref{eq:defWE} we have that the bound in \cref{eq:elowboundgen} takes the form $\mathcal E^{\rm low}(\varrho , W)= \max\{0, -\aver{W}_\varrho\}$ and is immediately obtained by taking any witness $W \in \mathcal{M}$.
Relevant examples of these measures, which are also of the distance-like form \eqref{eq:distmeasgen} are the \textit{best separable approximation (BSA)}~\cite{LewSanp98,KarnasLew01}, and the {\it generalized robustness} (GR)~\cite{SteinerPRA03}. 
In the following, we focus on the BSA as a paradigmatic example for concreteness, but our methods can also be applied to the GR, which we discuss in \cref{app:GR}.
This is based on expressing a state as a convex mixture of a separable contribution and a (generally entangled) remainder~\cite{LewSanp98}:
\begin{equation}
\label{eq:LS_BSA_decomp}
\varrho = (1-t)\,\sigma + t\,\nu,
\qquad
\sigma\in\Sep,\ \nu\ge 0,\ \tr(\nu)=1,\ \ t\in[0,1].
\end{equation}
The BSA is then obtained by minimizing the weight $t$ of the remainder (equivalently, maximizing the separable weight $1-t$)~\cite{LewSanp98}:
\begin{equation}
\label{eq:BSA_def}
\BSA(\varrho):=\min\Bigl\{\,t\ge 0\ :\ \exists\,\sigma\in\Sep,\ \nu\ \text{s.t.}\ \varrho=(1-t)\sigma+t\nu\,\Bigr\}.
\end{equation}
This can be seen as the dual problem to \cref{eq:defWE} for the choice 
$\mathcal M_{\rm BSA}=\{W\in\EW | \id +W \geq 0 \}$~\cite{brandao05}, and also corresponds to the form in \cref{eq:distmeasgen} with 
\be\label{eq:DBSA}
D_{BSA}(\varrho , \sigma)= \min_{t \in [0,1]} \tr(\varrho - (1-t) \sigma),
\ee
subject to $t\nu = \varrho - (1-t)\sigma \geq 0$. 

Notably, $\mathcal{E}_{BSA}(\Psi) = 1$ for any pure entangled state $\ket{\Psi}$, the optimal witness being $W = \alpha \id - \ketbra{\Psi}$ where $\alpha$ is the largest Schmidt coefficient across any bipartition, corresponding to the maximal overlap of any fully separable state with $\ket{\Psi}$.
In our application, this implies that for non-degenerate ground states, the BSA is always one, and can be optimally quantified from the fidelity to the ground state itself. However, this does not hold for thermal mixtures of degenerate ground states, especially if some are separable. Still, evaluating such a fidelity, for example for low-temperature states would be quite difficult in a many-body scenario, where the ground state is not fully characterized.
In general, the BSA is highly complex to evaluate exactly for mixed states (as any entanglement monotone). In fact, it is also a hard numerical problem even for two or three qubit states~\cite{LewSanp98,GabdulinMandilara2019}, and it becomes quickly intractable already for few-qubit systems~\cite{GabdulinMandilara2019,GirardinBrunnerKrivachy2022}.

Crucially, for our purposes, $\BSA(\varrho)$ admits lower bounds from any normalized witness, namely
any entanglement witness such that
\begin{equation}
\label{eq:witness_normalization_BSA}
W\ \text{is a witness},\qquad W\ge -\id ,
\end{equation}
which is due to the dual definition mentioned earlier and can be also seen as follows. For any BSA-type decomposition \cref{eq:LS_BSA_decomp} we have
\begin{equation}
\tr(W\varrho)=(1-t)\tr(W\sigma)+t\tr(W\nu)\ \ge\ (1-t)\cdot 0 + t\cdot(-1)=-t,
\end{equation}
where we used $\tr(W\sigma)\ge 0$ for all $\sigma\in\Sep$ and $\tr(W\nu)\ge -\tr(\nu)=-1$ from $W\ge -\id$.
Therefore,
\begin{equation}
\label{eq:BSA_witness_bound}
\BSA(\varrho)\ \ge\ \max\{0,\ -\tr(W\varrho)\}
\qquad\text{for all witnesses }W\text{ with }W\ge -\id ,
\end{equation}
which characterizes the BSA as a monotone in the class \eqref{eq:defWE} with the set $\mathcal{M}_{BSA}$ that we mentioned before.
In this sense, $\BSA(\varrho)$ is simultaneously (i) a distance-like entanglement monotone defined by an optimal convex decomposition and (ii) a special case of a witness-based entanglement monotone, defined from a set $\mathcal M_{\rm BSA}$ such that any witness can be normalized to belong to it~\cite{brandao05,FadelVitagliano_2021,CramerEtAl2013}.

\subsection{Spin-squeezing criteria and entanglement}
\label{sec:spin_squeezing_section}

Spin-squeezing emerged historically in the context of quantum optics, atomic physics and quantum metrology, as a concept aimed at surpassing classical limits on precision measurements imposed by uncorrelated particles~\cite{Ma2011Quantum}. Contrary to squeezing of bosonic quadratures, spin-squeezing has a multi-faceted definition, due to the different algebraic structure of the spin operators with respect to two canonical quadratures. This notion is also useful in many-body physics in the theory of collective spin systems, which includes paradigmatic models such as those introduced by Dicke~\cite{Dicke1954Coherence,KirtonRosesKeelingDallaTorre2018DickeReview} and Lipkin-Meshkov-Glick~\cite{Lipkin1965188,Meshkov1965199,Glick1965211,Ribeiro_2007}. If we consider ensembles of two-level systems described by collective spin operators, we can identify classical states of these ensembles as spin-coherent states, which provide a natural many-body analog of Glauber coherent states. Metrologically speaking, spin coherent states represent the benchmark for what is possible with separable states (so-called standard quantum limit (SQL)). 
The most well-known definition of a spin-squeezing parameter was introduced in Refs.~\cite{KitagawaUeda1993, WinelandBollingerItanoMooreHeinzen1992, WinelandPRA1994} and captures  the metrological gain that can be obtained when using a correlated state versus an uncorrelated state, i.e., a spin-squeezed versus a spin-coherent state.
Improving upon this limit motivated the research for collective states with reduced noise in one or two spin directions at the expense of increased noise in a conjugate direction. Such spin-squeezed states arise naturally from nonlinear collective interactions such as one-axis twisting and two-axis counter-twisting \cite{KitagawaUeda1993,Ma2011Quantum}. 

Based on these metrological findings, it soon became apparent that these spin-squeezed states are deeply connected to many-body entanglement, and that these spin-squeezing parameters are in fact also entanglement witnesses \cite{SoerensenMoelmer2001,SoerensenMoelmer2001}. While early works focused on a single optimized direction perpendicular to the mean spin, subsequent works generalized this idea to a broad class of inequalities involving the first and second moments of collective spin operators along arbitrary orthogonal directions. This led to the emergence of the generalized set of spin-squeezing inequalities \cite{tothPRL07,tothPRA09} which was later generalized to higher spins \cite{vitagliano11,vitagliano14,Vitagliano2025sudsqueezingmany}. 

Formally, let us now define the collective spin operators $J_k = \sum_{n=1}^N j_k^{(n)}$, where $j_k = \tfrac 1 2 \sigma_k$ with $k \in \{x,y,z\}$ are the single-particle spin directions.
For such an ensemble, the generalized SSIs of Refs.~\cite{tothPRL07,tothPRA09} read:
\be\label{eq:GenSSIsFull}
\begin{aligned}
\aver{J_x^2}+\aver{J_y^2}+\aver{J_z^2} &\leq N(N+2)/4 , \\
(\Delta  J_x)^2+(\Delta  J_y)^2+(\Delta  J_z)^2 &\geq N/2 , \\
(N-1)(\Delta  J_k)^2-\aver{J_l^2}-\aver{J_m^2} &\geq -N/2 , \\
(N-1)[(\Delta  J_k)^2+(\Delta  J_l)^2]-\aver{J_m^2} 
&\geq N(N-2)/4,
\end{aligned}
\ee
plus all permutations of directions $(k,l,m) \in \{x,y,z\}$. Notably, this set of inequalities allows one to detect entanglement in a wide range of many-body settings, including Dicke states, thermal states, and in general states generated by dissipative or driven dynamics, especially in the context of spin ensembles interacting collectively~\cite{Ma2011Quantum}.
Moreover, \cref{eq:GenSSIsFull} becomes a complete set of entanglement conditions in the limit of large $N$, in the sense that if a state is not detected by any of these inequalities, then there is a corresponding separable state with the same values of the first and second collective spin moments~\cite{tothPRL07,tothPRA09}.

This is because the inequalities in \cref{eq:GenSSIsFull} define a polytope in the space of collective first and second moments, that is fully filled by separable states.
The extremal points of such a polytope correspond, in the limit $N\gg 1$, to separable permutationally invariant states of either of the following two types:
\begin{subequations}\label{eq:extremalsepstates}
\begin{align}
\varrho_{\rm s} &= p \updm^{\otimes N}_z + (1-p) \downdm^{\otimes N}_z , \\
\varrho_{\rm p} &= \mathcal{T}_{\rm PI}\!\left(\updm^{\otimes M}_z \otimes \downdm_z^{\otimes (N-M)} \right) ,
\end{align}
\end{subequations}
and similar states pointing in other directions on the sphere. Note that the top states are symmetric, while the bottom ones are just permutationally invariant. 
Note also that the states $\varrho_{\rm p}$ are constructed precisely via permutationally invariant twirling from product states, that can be seen as generalizations of spin-coherent states spanning all spin-$j$ subspaces with $0\leq j \leq N/2$. Here, $M$ is an integer that depends on the total spin length $|\aver{\vec J}|$. In the particular case $\aver{\vec J} = (0,0,0)$ (i.e., an unpolarized state) we have $M=N/2$. 
Thus, up to an $O(1/N)$ correction the whole polytope can be filled with separable states, obtained by a suitable mixture of extremal states. This
means that in the limit $N\rightarrow \infty$ no other inequalities than \cref{eq:GenSSIsFull}, based on first and second collective spin moments can detect states not detected by this set.

A useful aspect of this geometric characterization is that these ``generalized coherent'' boundary points need not correspond to the traditional picture of a fully symmetric, fully polarized spin-coherent state in the $j=N/2$ subspace. Rather, they should be understood as \emph{extremal separable configurations} compatible with given collective first and second moments: they are still product states (hence separable), but their microscopic structure may populate different total-spin sectors. In this sense, the SSI polytope provides a more general notion of ``classical reference'' than the usual spin-coherent benchmark.

It is worth noting that the collective second moments appearing in the SSIs can be calculated from average two-body correlations, which can be calculated in many cases, and often even exactly via a mean-field ansatz in the thermodynamic limit~\cite{HaukeHeylTagliacozzoZoller16, Wiesniak05, marty14}. Because of this, the SSIs can be related to entanglement measures between two spins, e.g.\ the concurrence~\cite{VidalConc2006,amico08,Ma2011Quantum}. Nevertheless, it has been shown that the complete set of SSIs in \cref{eq:GenSSIsFull} can detect entanglement even if the two-particle states are separable~\cite{tothPRL07,tothPRA09,Vitagliano2025sudsqueezingmany}. This can be understood as a marginal problem: even if the two-body reduced density matrices are themselves separable, they are not compatible with a global separable permutationally invariant state, which in turn can be also understood in terms of De Finetti arguments (that is, the two-body marginals of any permutationally invariant state, also entangled, become separable in the thermodynamic limit $N\rightarrow \infty$)~\cite{STORMER196948,Hudson1976,Fannes1988,CavesFuchsSchack2002,KonigRenner2005,Christandl2007,KonigMitchison09,Trimborn_2016}.

As a final remark, going beyond first and second moments, other criteria based on collective spin operators, and in particular also criteria in terms of third-order moments of such operators have been derived in \cite{KorbiczCiracLewenstein2005,KorbiczEtAl2006}. These witnesses can potentially detect states not detected by the generalized SSIs; however, they detect only states that have a three-particle marginal that is not PPT, and require quantities more complex to extract in many-body states. 
\subsection{Finding the closest separable state from an iterative ansatz}\label{sec:ansatzintro}

Beyond witnesses, to fully certify the  distance to the set of fully separable states, {\it upper bounds} are needed. This problem is similarly (if not more) complex, as it requires an optimization over quantum states or decompositions~\cite{NavascuesPRL2009,barreiro2010experimental,KayPRA2011,GUHNE2011406,KampermannPRA1012,brierley2017convexseparationconvexoptimization,ShangGuhne2018,TiesScipost2024}. At the same time, a generic upper bound can be calculated by finding the distance to any given separable state, which can then be minimized over all (or a subset of) separable states. Numerically, this can be attempted using convex optimization algorithms~\cite{GuhneReimpellWernerPRL07,KampermannPRA1012,ShangGuhne2018}, where the feasible set is the convex hull of pure product states, or symmetrizations thereof, as discussed in \cref{sec:ent_dist_monotones}.

\begin{figure}
    \centering
    \includegraphics[width=.6\linewidth]{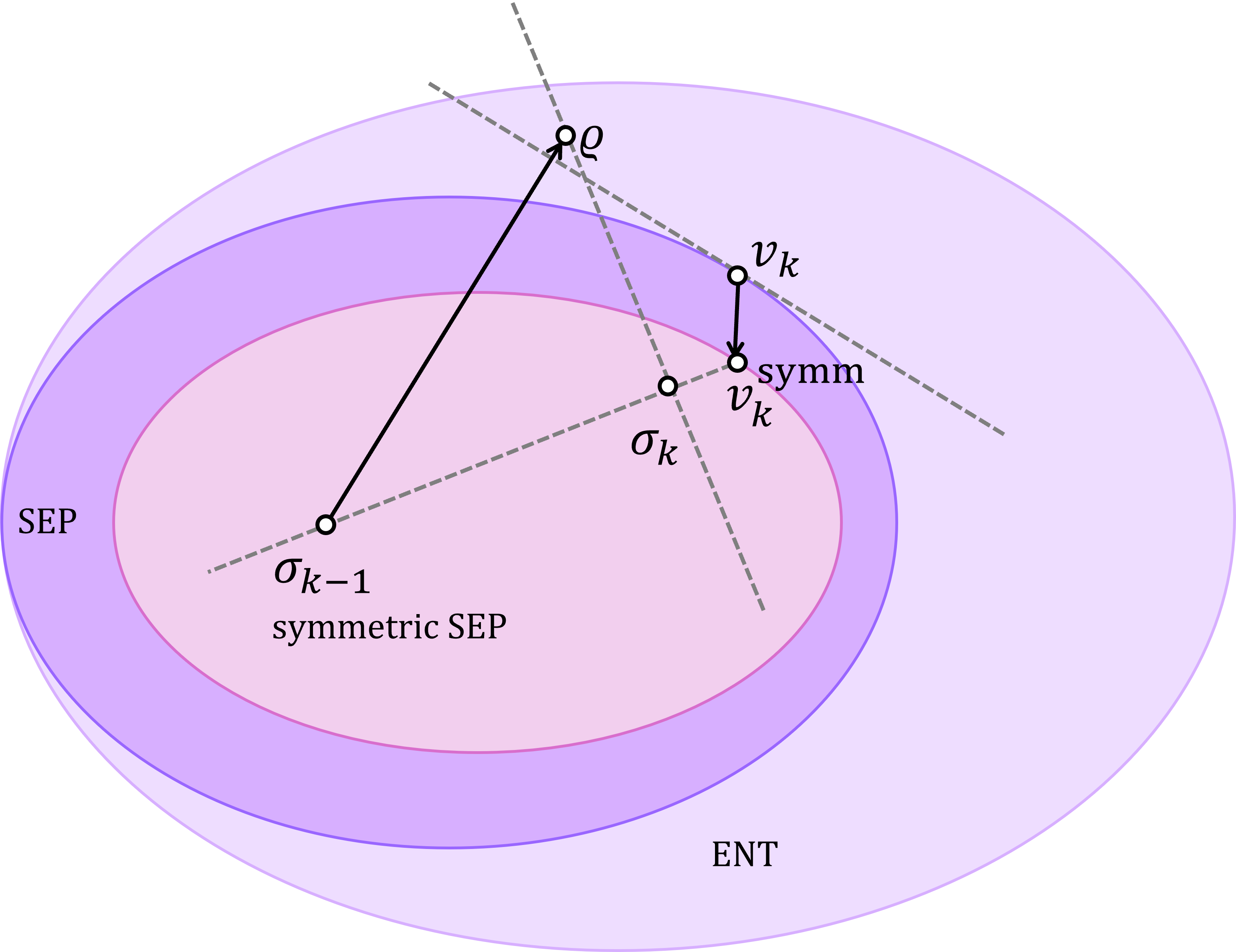}
    \caption{At step $k$, starting from a separable state $\sigma_{k-1}$, we find a pure product state with maximal overlap ($v_k$) and symmetrize it ($v_k^{\rm symm}$). Then we update our separable ansatz as the state that minimizes the distance to the separable set. For the first iteration, we can initialize $\sigma_0$ with, e.g., the fully polarized state.
    }
    \label{fig:algo}
\end{figure}

Let us illustrate this method to find an upper bound that focuses on the BSA, but we note that similar methods can be applied to other measures of the form \eqref{eq:distmeasgen}. To find a tight upper bound to the BSA, we must find the decomposition of the form as in \cref{eq:DBSA} with the ``closest'' possible $\sigma \in {\rm SEP}$, which can be done with a method similar to Ref.~\cite{ShangGuhne2018} and improved by exploiting the symmetry of the state. Concretely, at iteration $k$ we consider a separable ansatz of the form 

\begin{equation}
    \sigma_k = \sum_{i = 1}^{m_k} p_i^{(k)} \ketbra{v_i^{(k)}},
\end{equation}

where each $\ket{v_i^{(k)}} = \ket{\psi_1 \dots \psi_N}$ is a fully separable pure product state and the $p_i^{(k)}$ form a probability distribution. In the following steps (cf. \cref{fig:algo}), we explain how to obtain an iteratively better separable ansatz state at each step: 

\begin{enumerate}
    \item Start with some separable state $\sigma_{k-1}$as input, e.g., initially it could be the totally mixed state $\sigma_0 = \id/ 2^N$), or a fully polarized state.
    \item Choose a new product state $ \ket{v_i^{(k)}} = \ket{v}$ by considering the overlap optimization problem \footnote{Note that the $\lambda \approx O(1)$ only ensures positivity and numerical stability.},
    $$\max_{\ket{v}} \abs{\bra{v} (\varrho - \sigma_{k-1} + \lambda \mathbb{1}) \ket{v}}.$$
    As this is the key step, which is typically very complex, the practical goal is not to find the absolute optimal solution, but to find a good approximate solution in a feasible way. To illustrate it for a two-particle state, the idea is to fix one party $\ket{a}$ and optimize over $\ket{b}$ using the procedure described in \cite{KampermannPRA1012} (see also \cite{GuhneReimpellWernerPRL07,GuhneReimpellWerner08}), i.e., 
    $$\max_{\ket{b}} \ \bra{b} \bra{a}(\varrho - \sigma_{k-1} + \lambda \mathbb{1}) \ket{a} \ket{b}$$
    The optimal $\ket{b}$ is then the eigenvector of the largest eigenvalue of the matrix given by $\bra{a} (\varrho - \sigma_{k-1} + \lambda \mathbb{1}) \ket{a}$ (at fixed $\ket{a}$).
    Repeating this step and alternating between fixing $\ket{a}$ or $\ket{b}$ moves the state iteratively closer to the optimal solution by monotonically increasing the overlap. This scheme will in general only converge to a local maximum of the overlap function. This algorithm straightforwardly extends to multipartite states that are fully separable.
    If the states have some symmetry, such as permutational invariance, as we will explain, it is possible to simplify this procedure further. 
    \item Add the resulting $\ket{v_{i}^{(k)}}$ to the ensemble $\{\ket{v_{i}^{(0)}}, \ket{v_{i}^{(1)}}, ..., \ket{v_{i}^{(k-1)}}\}$ and update the probabilities by solving the optimization
        $$\min_{\{p_i > 0 \}}  \norm{\sigma_k - \varrho}^2,$$
    with $\sigma_k = \sum_i p_i^{(k)} \ketbra{v_i^{(k)}}$, now including the new product state. In our numerical studies, this step is solved using a sequential quadratic programming routine. 
    \item Convergence check: If the target state is very close to the optimal state or does not get any closer, stop the algorithm and output $\sigma = \sigma_k$. Otherwise, proceed with step 2.
    Formally, to ensure rigorously that the ansatz gets sufficiently close to the target (i.e., such that their difference is a separable state) one can rely on results on the so-called separable ball around the maximally mixed state~\cite{gurvitsbarnum2002,Braunsteinetal1999,zyczkowskietal1998,KampermannPRA1012}.
\end{enumerate}

Once the algorithm has stopped, compute the upper bound to the entanglement monotone $D(\rho, \sigma)$ corresponding to the final ansatz $\sigma$.

Note that instead of considering the two-norm distance in step 3, one can also consider a different distance measure, e.g., the one appearing in the definition of the concrete monotone. However, in all cases we considered, the two-norm distance performed best in terms of convergence speed and numerical stability. Overall, the iterative procedure described here provides a separable state that is closer and closer to the target state $\varrho$ and 
will eventually converge or can be terminated at will, e.g., when the distance is close to a given value depending on a given precision. 
As we mentioned, in practice when the difference between the separable ansatz and the target density matrix state becomes very close to the maximally mixed state, one can also show analytically that the state must be separable and thus terminate the algorithm. This is due to the existence of a {\it separable Ball} around the maximally mixed state~\cite{gurvitsbarnum2002,Braunsteinetal1999,zyczkowskietal1998,KampermannPRA1012}.

Note that other algorithms exist for looking for the separable state closest to a given state, e.g., based on mixing random product states via linear programming~\cite{GabdulinMandilara2019}, or with machine learning methods~\cite{GirardinBrunnerKrivachy2022}, the bottleneck always being the large number of product states in a generic separable decomposition, as well as more generally the extremely rapid growth of the number of relevant product states to consider.
In fact, currently, all algorithms can find the separable decomposition for generic states up to three qubits, and in special cases up to four qubits.

Thus, clearly in the most general case optimizing the upper bounds becomes extremely challenging, as the problem becomes computationally infeasible already for few-qubit systems. However, it could be potentially simplified considerably when the quantum states have some symmetry, as is also illustrated in \cref{fig:algo}. Formally, a state can be such that
\be\label{eq:twirlopmain}
\varrho = \sum_{U_i \in G} U_i \varrho U_i^\dagger := \mathcal P_G(\varrho),
\ee
where $G$ is a subgroup of unitaries that do not increase entanglement, and the sum could also be substituted by an integral for continuous symmetry groups. In this case, as we will discuss later, applying the appropriate twirling on the product state found after step $2$ can considerably speed up the algorithm.

\subsection{Permutation invariant XXZ model at equilibrium}
\label{sec:perm_inv_XXZ}
As we mentioned, spin-squeezed states arise quite naturally as an important class of states in the many-body setting, in the context of quadratic spin models with fully-connected interaction graphs. These models play a central role both in many-body physics and in quantum metrology, and they are frequently approached using semi-classical or mean-field techniques, where the many-body state is approximated by a spin-coherent (product) ansatz and the dynamics is reduced to an effective classical evolution on the collective Bloch sphere.
Paradigmatic models to investigate spin-squeezed and related collectively correlated states are the Lipkin--Meshkov--Glick (LMG)~\cite{Lipkin1965188,Meshkov1965199,Glick1965211} or the Dicke model~\cite{Dicke1954Coherence}, among others~\cite{Ma2011Quantum,KirtonRosesKeelingDallaTorre2018DickeReview,DEFENU20241}.

These models are well studied both at equilibrium and dynamically, including their quantum critical and dynamical phase structure~\cite{Ma2011Quantum,KirtonRosesKeelingDallaTorre2018DickeReview,DEFENU20241}, and in many regimes their behavior admits accurate semiclassical descriptions in the large-$N$ limit ~\cite{Ribeiro_2007,KirtonRosesKeelingDallaTorre2018DickeReview}.
Still, entanglement in such settings has also been investigated extensively, especially for ground states and for pure-state out-of-equilibrium protocols, where entanglement can be tracked as a function of time or system parameters~\cite{Latorre_2005,Vidal2004Entanglement_first,vidaldynamicsLMG04,OrusPRA2008}.
This has been done mainly via the scaling of entropies of bipartitions with the size of the subsystem, but also via geometric measures of entanglement in some cases, which fall into the same (or a similar) class of distance-based monotones as defined in \cref{eq:distmeasgen} ~\cite{WangMolmer2002,WangSanders2003,stockton03,LambertEmaryBrandesPRL2004,Vidal2004Entanglement_second,vidaldynamicsLMG04,Vidal2004Entanglement_first,Latorre_2005,LambertEmaryBrandesPRA2005,VidalConc2006,BarthelDusuelVidalPRL2006,Vidal_2007,OrusPRL2008,OrusPRA2008,OrusDusuelVidalPRL2008,Or_s_2010,wang10,yin11}.

Generically, we can write a broad range of such models as the following spin-$1/2$ Hamiltonian:
\be\label{eq:HXXZ}
H= \frac g N \left( J_x^2 + J_y^2 \right) + \frac{g_z} N J_z^2 + h J_z ,
\ee
which is essentially a fully-connected XXZ model. Based on the sign of $g$, we can distinguish two cases: {\it ferromagnetic} $(g<0)$ and {\it anti-ferromagnetic} $(g > 0)$.
In both cases there are different phases, depending on the values of $g_z$ and $h$.  In particular:
\begin{itemize}
    \item For $g,g_z>0$ (AFM) and $h=0$, the ground state is a many-body singlet, 
\be
\ket{\Psi_{\rm sing}} = \ket{J=0, J_z=0, i_J} = \ket{\psi_-}^{\otimes N/2},
\ee
with $\ket{\psi_-}= \tfrac 1 {\sqrt 2} (\ket{\hspace{-3pt}\uparrow \downarrow}_z - \ket{\hspace{-3pt}\downarrow \uparrow}_z)$ being a product of two-particle singlets. The ground state subspace has degeneracy given by $\mu_{0} = \binom{N}{N/2} - \binom{N}{N/2 - 1}$.
\end{itemize}
For $g_z=0$ we get an XX model, which has itself different phases with different ground states, depending on $h$:
\begin{itemize}
    \item For $g<0$ (FM) we have:   
    \begin{itemize}
        \item When $h=0$ the (non-degenerate) ground state is the symmetric Dicke state with $N/2$ excitations, $\ket{J=N/2, J_z=0, i_J=1}$. 
        \item  When $h > g$ the ground state is fully polarized, given by the state $\ket{N/2, N/2}$, which is a pure product state (the direction depending on the sign of $h$).
    \end{itemize}
    In particular, a second-order QPT arises in the thermodynamic limit $N\rightarrow \infty$ for $h\rightarrow g$.
    \item For $g>0$ (AFM) we have:    
    \begin{itemize}
        \item For $h\geq g/N$ the ground state is again the fully-polarized state $\ket{N/2, N/2}$ and thus a product state. 
        \item For $h < g/N$, the ground-state space is very degenerate as there is a ground state in each spin-$J$ sector which is given by $\ket{J,J}$, which is a consequence of {\it supersymmetry}~\cite{Vidal2004Entanglement_first}. This includes product states (e.g., for $J=N/2$) and entangled states (e.g., the singlet states for $J=0$). 
    \end{itemize}
    In particular, a first-order QPT arises at $h\rightarrow 0$ (again in the limit $N\rightarrow \infty$).
\end{itemize}

Ground state entanglement has been studied quite extensively across these phases~\cite{LambertEmaryBrandesPRL2004,Vidal2004Entanglement_second,vidaldynamicsLMG04,Vidal2004Entanglement_first,Latorre_2005,LambertEmaryBrandesPRA2005,VidalConc2006,BarthelDusuelVidalPRL2006,Vidal_2007,OrusDusuelVidalPRL2008}.
In practice, as mentioned, such fully connected models are most often analyzed using large-$N$ mean-field methods, where the many-body state is approximated by a spin-coherent ansatz and quantum fluctuations are incorporated perturbatively, e.g. via Holstein-Primakoff or Bogoliubov expansions around a classical trajectory \cite{Ribeiro_2007, Vidal2004Entanglement_first, KirtonRosesKeelingDallaTorre2018DickeReview}. Remarkably, such semiclassical treatments are sufficient to compute two-body correlators and thus the second moments of collective spin operators. Since the SSIs depend only on these collective moments, they can be evaluated with the same methods. Additionally, the two-spin reduced density matrices of symmetric states can be obtained explicitly, allowing for the evaluation of bipartite entanglement measures such as the concurrence~\cite{VidalConc2006}. In the context of the LMG and related models, the concurrence has been computed both for ground states and dynamical protocols, and shown to be closely connected to the original Kitagawa-Ueda spin-squeezing parameter \cite{WangMolmer2002, WangSanders2003, stockton03, LambertEmaryBrandesPRL2004, Vidal2004Entanglement_second}. In particular, for symmetric states, the presence of spin squeezing implies nonzero pairwise concurrence, although the converse does not hold in general.

Going beyond ground states, we are interested in finite temperature states, in particular thermal states with a density matrix given by the Gibbs ensemble, $\varrho_T = \frac 1 {\mathcal Z_T} e^{-H/T}$~\footnote{We set the Boltzmann constant as well as the reduced Planck constant to $k_B = \hbar =1$.}, where the partition function is
\be\label{eq:partition_function}
\mathcal Z_T= \sum_{J=0}^{N/2}  \mu_J \sum_{M_z=-J}^{J} e^{- (g J(J+1)/N - (g -g_z) M_z^2 + h N M_z)/NT} .
\ee
We emphasize that all spin $J$ subspaces with $0\leq J \leq N/2$ must be taken into account. States of this form are closely related to spin-squeezed states and, as we mentioned, several witnesses have been developed targeted to detect entanglement in those states. Furthermore, these witnesses have been shown to meaningfully bound entanglement monotones~\cite{stockton03,KrammerPRL09,CramerPlenioWunderlich11,marty14,jing2019,FadelVitagliano_2021}.
Thus, it is natural to think that SSIs provide optimal entanglement witnesses, and then potentially tight lower bounds to entanglement monotones for such thermal states. 

In the following, we discuss this question in a precise way, making use of all the tools discussed earlier.

\section{Results and Discussion}\label{sec:results}

In this section, we discuss our approach to estimate entanglement in collective spin systems beyond pure states. Building upon what we described earlier, we focus on states that arise in the context of spin-squeezing and related spin models, trying to characterize entanglement in
equilibrium states as a paradigmatic application that allows us to discuss some key features of our approach and also explore an interesting physical scenario, with a potentially broad application.

We also stress that there is growing interest in collectively correlated spin states beyond the typical restriction of the total spin sector $j=N/2$, which is what we aim at investigating. 
Prominent examples are singlets and, more generally, low-$j$ sectors, which naturally arise in AFM or frustration-dominated scenarios and can be tied to a richer landscape of quantum phases and correlations than in FM-like, symmetric-sector physics. Such singlet-like macroscopic states have also become experimentally relevant in cold-atom platforms~\cite{behbood14,mann2025squeezingclassicalantiferromagnetsquantum}.

\subsection{Finding a lower bound via an optimal spin-squeezing parameter}
\label{subsec:SSIs_BSA}

A central message of the previous sections is that for witnessing entanglement in collective spin states, especially with an experimental implementability motivation, SSIs have been derived~\cite{SoerensenNAT2001,KorbiczCiracLewenstein2005,KorbiczEtAl2006,tothPRL07,tothPRA09,Ma2011Quantum}, which, once again essentially provide a complete quantitative distinction between spin-squeezed and spin-coherent states. 
Importantly, SSIs are \emph{nonlinear} in the state: they are quadratic expressions in expectation values of spin operators onto quantum states.
At the same time, they may be understood as concise encodings of \emph{families of linear witnesses}, for given values of the average spin operators $\aver{\vec J}$.
Formally, this can be seen from the fact that the variance of an operator onto a quantum state can be written as~\cite{Dammeier_2015,FadelVitagliano_2021}:
\be\label{eq:linearized_variance}
(\Delta A)_\varrho^2 = \min_{a \in \mathbb R} \aver{\left( A- a \id \right)^2}_\varrho .
\ee
As we mentioned, among spin-squeezing-like inequalities, the ones introduced in Refs.~\cite{tothPRL07,tothPRA09} are particularly powerful, as they form a complete set of separability conditions within the information content of first and second collective moments. Moreover, they have a geometric description that, as we are going to observe, can be translated from the space of second collective spin moments, to the space of density matrices, 
to bound the distance of a given state to the set of fully separable ones.

In the following we (i) rewrite the set of generalized SSIs in a compact form, and (ii) translate its violation into a normalized witness in $\mathcal M_{\rm BSA}$, yielding an easily computable lower bound on the best separable approximation.
Crucially, the resulting quantity implicitly accounts for the full SSI family at once, rather than selecting a witness case-by-case.

First, for point (i), let us rewrite the complete set of SSIs in \cref{eq:GenSSIsFull} in a rotation-invariant form~\cite{tothPRL07,tothPRA09} (see also \cite{vitagliano14,Vitagliano2025sudsqueezingmany}).
Let us consider the following matrices of second moments of the collective spin operators:
\be\label{eq:coll_matrices}
\begin{aligned}
    (C_\varrho)_{kl}&= \tfrac 1 2 \langle J_k J_l +J_l J_k\rangle_\varrho , \\
(\Gamma_\varrho)_{kl}&:= (C_\varrho)_{kl} 
- \langle J_k\rangle_\varrho \langle J_l\rangle_\varrho , \\
(\mathfrak X_\varrho)_{kl} &:= (\Gamma_\varrho)_{kl}
+ \tfrac 1 {2(N-1)} \langle J_k J_l + J_l J_k\rangle_\varrho
- \tfrac{N^2}{4(N-1)} \delta_{kl} .
\end{aligned}
\ee
The complete set of SSIs in \cref{eq:GenSSIsFull} can be cast in a rotation-invariant form, in terms of $\tr(\Gamma_\varrho)$ and (a subset of) the eigenvalues of $\mathfrak X_\varrho$, which we denote by $\lambda_k(\mathfrak X_\varrho)$.
This reads:
\be\label{eq:GenSSIsFull_rotinv}
\begin{gathered}
\tr(C_\varrho) \leq N(N+2)/4 , \\
\tr(\Gamma_\varrho) \geq N/2 , \\
\tr(\Gamma_\varrho) - N/2 - \lambda_k(\mathfrak X_\varrho) - \lambda_l(\mathfrak X_\varrho) \geq 0 , \\
\tr(\Gamma_\varrho) - N/2 - \lambda_k(\mathfrak X_\varrho) \geq 0 , 
\end{gathered}
\ee
which can be seen from simple algebra, by noting that the diagonal elements of $\mathfrak X_\varrho$ can be expressed directly in terms of collective spin variances and second moments as
\be
(\mathfrak X_\varrho)_{kk}
= (\Delta J_k)_\varrho^2 + \tfrac 1 {N-1}\langle J_k^2\rangle_\varrho
- \frac{N^2}{4(N-1)} .
\ee
To reach the rotation-invariant form in \cref{eq:GenSSIsFull_rotinv} one considers optimal directions $k$ given by the eigenvalues of $\mathfrak X_\varrho$ as the principal axes $\{x,y,z\}$.
Moreover, here $\lambda_k(\mathfrak X_\varrho)$ and $\lambda_l(\mathfrak X_\varrho)$ denote arbitrary two eigenvalues of $\mathfrak X_\varrho$. 
Clearly, from \cref{eq:GenSSIsFull_rotinv} one can see that it is always better to take them as large (and positive) as possible, in order to observe a violation and thus detect entanglement.

From \cref{eq:GenSSIsFull_rotinv}, we can thus see that, denoting by $\lambda_k^{\rm pos}(\mathfrak X_\varrho)$ the positive eigenvalues of $\mathfrak X_\varrho$, the inequality which is optimal for a given state $\varrho$ can be written as a single \emph{spin-squeezing parameter} as follows (see also \cite{Vitagliano2025sudsqueezingmany} ):
\begin{gather}
\label{eq:SSIspar}
\xi_{SS}(\varrho)
:= \tr(\Gamma_\varrho) - \sum_{k=1}^{K}\lambda_k^{\rm pos}(\mathfrak X_\varrho) - \frac{N}{2} ,
\end{gather}
where $K\in\{0,1,2\}$ is the number of positive eigenvalues of $\mathfrak X_\varrho$.

Now let us proceed with point (ii), namely to find a lower bound to the BSA based on $\xi_{SS}(\varrho)$. 
First, we note that, although this parameter is quadratic in expectation values, it can be conveniently expressed as a minimization over
linear expectation values due to \cref{eq:linearized_variance},
which expresses each variance as a minimization of a linear expectation value over a shifted operator.
Applying this to the generalized SSIs, and working in the principal spin basis (labelled by $k\in\{1,2,3\}$), one can find the following.

\paragraph{Observation (normalized SSI bound on the BSA).}
\emph{Let $\varrho$ be an $N$-spin state and let $\xi_{SS}(\varrho)$ be defined as in~\eqref{eq:SSIspar}, with $K\in\{0,1,2\}$ the number of positive eigenvalues of $\mathfrak X_\varrho$.
Then the following lower bound holds:}
\be
\label{eq:BSA_bound_from_xiSS_obs}
\mathcal E_{\rm BSA}(\varrho)\ \ge\ -\frac{1}{\mathcal N_K}\,\xi_{SS}(\varrho),
\ee
\emph{with}
\be
\label{eq:N_K_def_obs}
\mathcal{N}_K
= \frac N 2  - K\frac{N^2}{4(N-1)} + \frac{N(N+2)}{8(N-1)}\,K(K-1).
\ee

\emph{Proof (sketch).} Equation~\eqref{eq:SSIspar} provides a compact expression for the optimal inequality detecting the given state $\varrho$, i.e., it performs the optimization over the family in \cref{eq:GenSSIsFull}. To obtain a witness in $\mathcal M_{\rm BSA}$, we normalize the corresponding optimal inequality by (the negative of) its minimal possible value, which ensures the correct scaling required by the BSA construction.
The relevant (negative) minima are: $N/2$ for $K=0$ and $N(N-2)/4(N-1)$ for $K=1$ (both attained by the singlet state), and $N^2/4(N-1)$ for $K=2$ (attained by the symmetric Dicke state with $N/2$ excitations).\footnote{
The last inequality in the set~\eqref{eq:GenSSIsFull} is trivially satisfied by all states and corresponds to $K=3$, hence it does not enter the parameter.}
Carrying out this normalization yields~\eqref{eq:BSA_bound_from_xiSS_obs} with the state-independent prefactor~\eqref{eq:N_K_def_obs}. See \cref{app:boundfromSSIs} for more details.
\hfill$\square$

Note that for many relevant collective spin states, the spin-squeezing parameter in \cref{eq:SSIspar}, and thus the bound to the BSA is easy to calculate, as it involves only second-order collective
spin moments. Numerically, as we will show afterwards, the calculation of this bound in these thermal states is relatively straightforward:
the ingredients $\langle J_k\rangle_{\varrho_T}$ and $(\Delta J_k)_{\varrho_T}^2$ entering $\Gamma_{\varrho_T}$ and $\mathfrak X_{\varrho_T}$ can be obtained from derivatives of the partition function.
Operationally, Eq.~\eqref{eq:BSA_bound_from_xiSS_obs} should be read as follows:
rather than selecting one particular spin-squeezing witness, we evaluate the single scalar quantity $\xi_{SS}(\varrho_T)$, which already incorporates the optimization over the full generalized-SSI family accessible from first and second collective moments, and then rescale it to match the BSA witness constraints.

\subsection{Finding an upper bound via a symmetry-improved iterative ansatz}
\label{sec:upper_bound_symm}
While the previous paragraph provides easily computable \emph{lower} bounds from collective moments, an \emph{upper} bound on $\mathcal E_{\rm BSA}$ requires an explicit separable ansatz state $\sigma$ and the evaluation of the corresponding value of the objective (cf.~Eq.~\eqref{eq:DBSA}).
If one has thus access to an algorithm that iteratively improves a separable candidate state so as to approach $\varrho$, then at each iteration one obtains a valid separable $\sigma$ and hence a valid upper bound on $\mathcal E_{\rm BSA}$.

In general, to find an upper bound, we follow two interconnected strategies, both exploiting the symmetry and the underlying structure of our problem: 
\begin{itemize}
    \item[(i)] We consider simple separable ansatz states that are built from the extremal states of the spin-squeezing polytope, as in \cref{eq:extremalsepstates}. 
    \item[(ii)] We build on the iterative strategy introduced in \cref{sec:ansatzintro}, but we crucially \emph{upgrade} it by enforcing the symmetries of the physical problem.
\end{itemize}

In the systems considered here, the relevant target states are invariant under particle permutations and under global rotations about the $z$ axis.
Hence, we restrict the separable search to states sharing these symmetries.
Any such state can be obtained by twirling a non-symmetric separable state via the symmetrizing map $\mathcal P_{\text{PI},z}(\varrho) = \mathcal P_z \circ \mathcal{T}_{\rm PI}(\varrho)$, where $\mathcal{T}_{\rm PI}(\varrho)$ is the PI twirling defined in \cref{eq:PI_twirl_map} and 
\be
\mathcal P_z(\varrho) = \int_{0}^{2\pi} \de \theta \ e^{-i\theta J_z} \varrho \ e^{i\theta J_z}.
\ee
More in general, each state with these symmetries can be written as (see \cref{app:separablePIstates})
\be
\label{eq:fullansatz}
\varrho_{\text{PI},z} = \sum_{J, J_z} \alpha_{J , J_z}\, \ketbra{J,J_z} \otimes \id_{\mu_J},
\ee
where $\ket{J,J_z,i_J}$ are common eigenstates of $J^2:=J_x^2+J_y^2+J_z^2$ and $J_z$, and $i_J\in\{1,\dots,\mu_J\}$ labels the multiplicity of the permutation irrep.
In particular, the entire state is characterized by the coefficients $\{\alpha_{J,J_z}\}$.
For a pure product state $\ket{\psi_1}\otimes\cdots\otimes\ket{\psi_N}$, these coefficients are
\be
\label{eq:alphaJJzmain}
\alpha_{J, J_z} = \sum_{i_J = 1}^{\mu_J} \abs{\langle \psi_1 \dots \psi_N \ket{J, J_z, i_J}}^2,
\ee
which leads to a significant reduction of complexity, even though the calculation of these coefficients becomes also computationally hard for larger and larger $N$ \cite{Kirby_2018}.
For the first strategy, we considered a low-complexity separable family tailored to the symmetries, namely
\be
\label{eq:simpleans}
\sigma(\vec p)=\mathcal P_{\text{PI},z}\!\left(
\sum_{M,\theta_i} p_{M,i}\;
U_{\theta_i}\Big(\updm_z^{\otimes M}\otimes\downdm_z^{\otimes (N-M)}\Big)U_{\theta_i}^\dagger
\right),
\ee
where the $U_{\theta_i}$ are global rotations about the $J_y$ axis and $\vec p=\{p_{M,i}\}$ is a probability vector optimized to minimize the resulting upper bound. Once again, this family can be interpreted as a mixture of states \eqref{eq:extremalsepstates}, that are at the vertices of the spin-squeezing polytope (up to a correction negligible for large $N$). 

Here, the twirling $\mathcal P_{\text{PI},z}$ is performed virtually, in the sense that it enforces invariance under permutations and global rotations in the $(x,y)$-plane without loss of generality, and it allows one to characterize $\sigma(\vec p)$ efficiently through the reduced data $\{\alpha_{J,J_z}\}$ instead of the full density matrix.
For the second strategy, we start from the baseline algorithm discussed in \cref{sec:ansatzintro}. This proceeds by alternating two main tasks:
(i) for the current iterate, identify a ``best'' \emph{pure product state} direction $\ket{v} = \ket{\phi_1 , \dots , \phi_N}$, and
(ii) update the separable candidate by mixing in that product state with optimal probability weights $\{p_i\}$.
In our implementation, step (i) is carried out by maximizing the usual overlap functional
\be
\label{eq:prod_overlap}
\max_{\ket{v}}\ \langle v| \varrho| v \rangle ,
\ee
as in the original algorithm we build upon. This optimization is performed in a see-saw way, by maximizing the overlap fixing $(N-1)$ particle states iteratively with different subsets of particles and taking the largest eigenvector of the corresponding marginalized matrix. 

It is also worth stressing that, although the BSA optimization defines its own objective, this overlap maximization provides a computationally efficient alternative that still improves iteratively the separable approximation in practice. Note in fact, that the optimum of the BSA objective function is often difficult or becomes numerically ill-defined~\cite{LewSanp98,KarnasLew01}.
In generic mixed states, however, such product-state updates tend to drift away from the symmetry sector of the target $\varrho$, which slows convergence.
Hence we insert an additional step:
after each iteration, we twirl the newly found product state into a mixed but symmetric separable state: 
\begin{itemize}
        \item[2.1] In the case of optimizing over some symmetry group, bring this state into its symmetrized form. Add this state to the ensemble. 
\end{itemize}
Concretely, given a product state $\ket{v}$ produced by~\eqref{eq:prod_overlap}, we apply the twirling $
\mathcal P_{\text{PI},z}(\ketbra{v}_k)$
over particle permutations and global rotations about the $z$ axis.
The resulting state is (i) still separable, (ii) lies in the same $(\text{PI},z)$ symmetry sector as the target, and (iii) can be represented efficiently via the coefficients~\eqref{eq:alphaJJzmain}.
Updating the candidate separable approximation using these \emph{twirled} states keeps every iterate inside the correct symmetry-reduced manifold.

As another speed-up, it is often important to choose smartly the initial ansatz $\sigma_0$.
Often we initialize the iteration with a thermal state at a temperature where separability can be certified with high accuracy (e.g., the highest temperature at which the lower bound becomes zero provides a natural starting point).
Once the algorithm has converged at that temperature, the resulting separable state can be used as a robust ansatz to evaluate Eq.~\eqref{eq:DBSA} also at neighbouring temperatures.

In general we observe that this symmetry-enforced update accelerates convergence substantially, and the improvement is particularly pronounced whenever $\varrho$ is actually separable: in that regime, the iterates often approach a bona fide separable decomposition rather than merely yielding a loose upper bound.
At the same time, a systematic understanding of the optimality of this strategy is still missing.
In particular, it remains an open question whether one can design an even faster symmetry-adapted scheme.
We leave these algorithmic refinements for future work.

In a similar way, one may also consider augmenting this algorithm with some initial memory that could either contain one or more ``educated-guess-ansatz-states'' or an ansatz state taken from the simpler approach. In practice, we considered these approaches but found them to be of limited usefulness as they essentially only speed up convergence at the very beginning. The final convergence speed remains roughly the same, as is the overall state found with or without such memory.

Concretely, the algorithm is numerically implemented as discussed in Algorithm~\ref{alg:bsa_upper_bound}.

\begin{tcolorbox}[colback=white,colframe=DarkRed, title= \textbf{ \large Algorithm: Upper bound to the BSA}]

Algorithm used for the numerics in this paper, given a computational-basis state $\varrho$ $(2^N \times 2^N)$, and the symmetry transform matrix $U$ $(2^N \times 2^N)$ (here the Schur transform corresponding to the coefficients in \cref{eq:alphaJJzmain}). The overlap optimization step is based on the github implementation of \cite{ShangGuhne2018}.

\begin{algorithm}[H]
    \begin{algorithmic}

    \Require Target state $\varrho$, symmetry transformation $U$, tolerance $\varepsilon$, maximum iterations $i_{\max}$
    \Ensure Separable approximation $\sigma$ and upper bound $D_{\rm opt}$
    
    \State \textbf{Initialization:}
    \State Initial separable state $\sigma_0$ is fully polarized

    \State \text{Symmetry reduction:} 
    \State $\varrho_{\mathrm{diag}} = \mathrm{diag}(U \varrho U^\dagger), \quad \sigma_{\mathrm{diag}} = \mathrm{diag}(U \sigma_0 U^\dagger)$
    \State $D = \|\varrho_{\mathrm{diag}} - \sigma_{\mathrm{diag}}\|^2$
    \State Initialize memory $M \gets \{\sigma_{\mathrm{diag}}\}$
    
    \For{$k = 1$ to $i_{\max}$}
    
      \State \textbf{Product-state optimization:}
      \State Choose a random product state $|v\rangle$
      \State $|v_{\mathrm{opt}}\rangle \gets 
             \arg\max_{|v\rangle} \langle v | (\varrho - \sigma) | v \rangle$
    
      \State $\sigma_v = |v_{\mathrm{opt}}\rangle\langle v_{\mathrm{opt}}|, \quad \sigma_{v,\mathrm{diag}} = \mathrm{diag}(U \sigma_v U^\dagger)$
      \State $M \gets M \cup \{\sigma_{v,\mathrm{diag}}\}$
    
      \State \textbf{Convex optimization:}
      \State Solve
      \[
         \min_{\{x_i \ge 0\}} 
         \left\| \sum_i x_i M_i - \varrho_{\mathrm{diag}} \right\|^2
      \]
      \State $\sigma_{\mathrm{diag}} = \sum_i x_i M_i$
    
      \State \textbf{Convergence check:}
      \State $D_{\mathrm{new}} =
             \|\varrho_{\mathrm{diag}} - \sigma_{\mathrm{diag}}\|^2$
      \If{$D_{\mathrm{new}} < D$} 
      \State $D = D_{\mathrm{new}}$
      \EndIf
      \If{$D_{\mathrm{new}} < \varepsilon$}
        \State $D_{\mathrm{opt}} = D$
        \State \textbf{break}
      \EndIf 
      \State \text{Back transformation:} $\sigma = U^\dagger \sigma_{\mathrm{diag}} U$
    \EndFor   
    \State \Return $\sigma_{\mathrm{diag}},D$
    \end{algorithmic}
    \caption{BSA upper bound}
    \label{alg:bsa_upper_bound}
\end{algorithm}
\end{tcolorbox}

In summary, the upper-bound computation can be viewed as a symmetry-adapted search for the closest separable approximation to $\varrho$. This search is extremely complex, especially when the state is entangled, and can be approached in different ways, that still contain several ``empirical'' steps, like the initialization and the selection of appropriate ensembles of product states (e.g., either via an iterative maximization of the overlap as in Step 2 of the algorithm, or via a more intuition-driven and simple ansatz as in the first strategy). Overall, by enforcing the symmetry via twirling, we substantially reduce the effective dimension of the problem and obtain markedly faster convergence, a feature that becomes visible in the XXZ numerical results presented below.

\subsection{Numerical results}
\label{sec:numerical_results}

\begin{figure}
    \centering
    \begin{subfigure}[t]{.6\textwidth}
        \includegraphics[width=\textwidth]{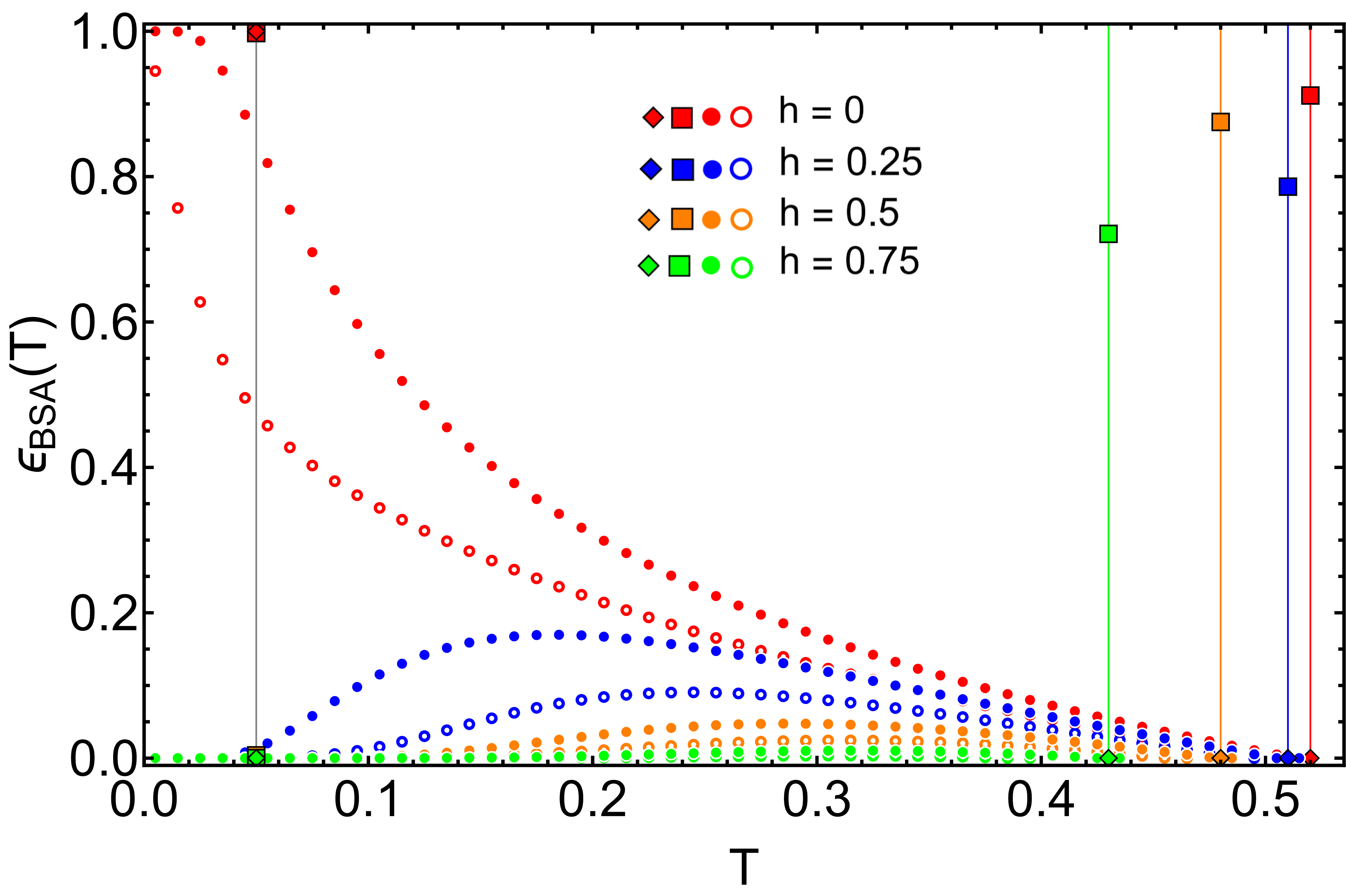}
        \caption{The FM case.}
        \label{fig:LMG_FM}
    \end{subfigure}
    \vfill
    \begin{subfigure}[t]{.6\textwidth}
        \includegraphics[width=\textwidth]{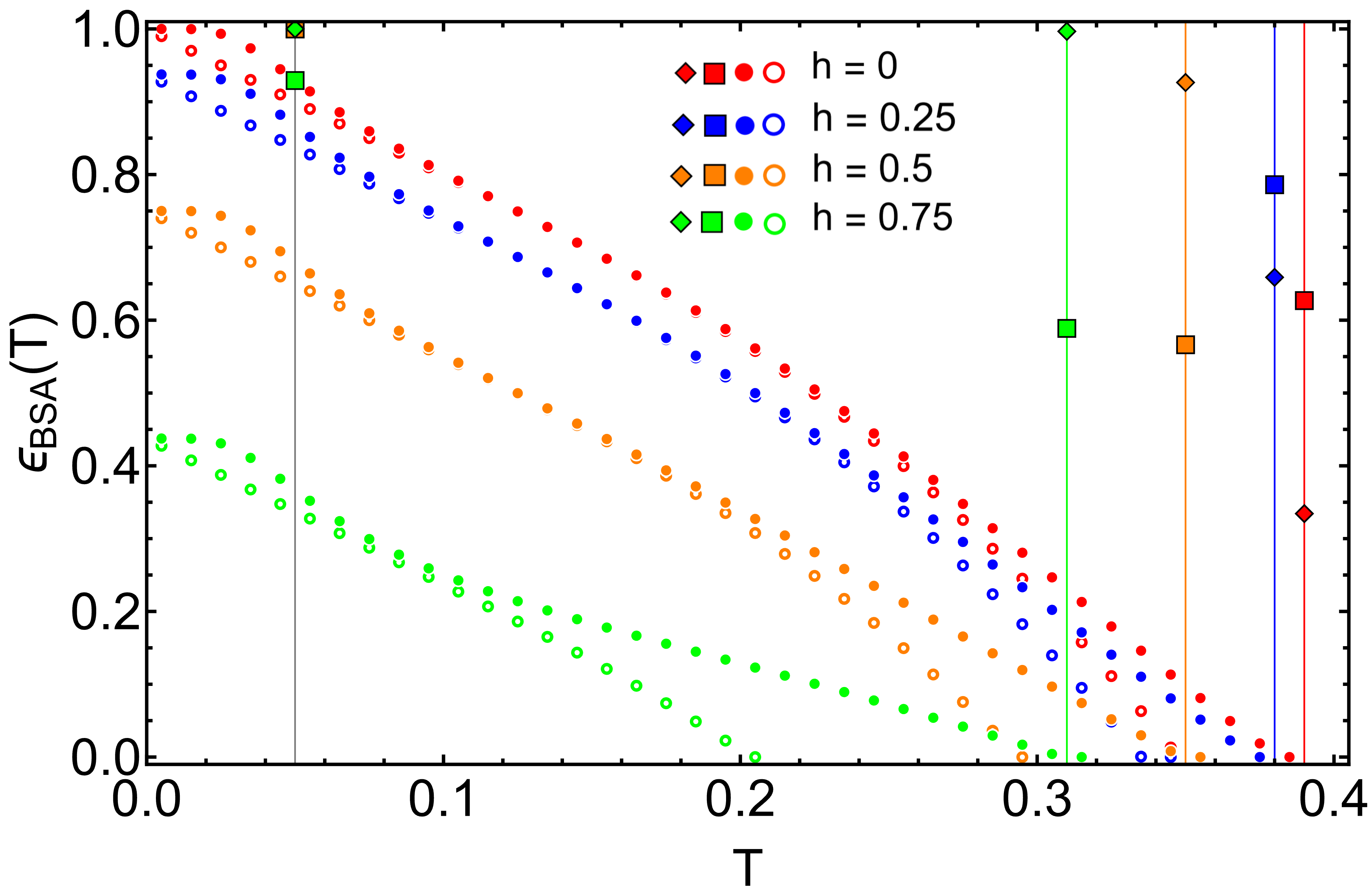}
         \caption{The AFM case.}
        \label{fig:LMG_AFM}
    \end{subfigure}
    \caption{Lower bounds to the BSA for $N = 200$ (circles) and for $N = 8$ (disks), upper bounds computed using the full optimization ansatz corresponding to Eq. (\ref{eq:fullansatz}) for $N = 8$ (diamonds) and from the simple ansatz in \cref{eq:simpleans} for $N = 8$ (squares). Vertical lines show the temperatures at which the respective upper bounds were computed.}
    \label{fig:LMG_lower_bounds}
\end{figure}

As we mentioned, to illustrate both the richness of the physics and the need for mixed-state methods, it is instructive to consider widely used collective models such as the fully-connected XXZ Hamiltonian, in both FM and AFM regimes. Overall, using these methods for the lower- and upper-bounds, we can make several observations. 
Firstly, for most ground states, the lower bound equals one, and is thus tight. Note that we consider $T\rightarrow 0$ thermal states, that are often highly mixed, and where it is not obvious that their BSA should be maximal. 
Secondly, the BSA becomes smaller than one in supersymmetric phases of the AFM XX model (cf. \cref{fig:LMG_AFM}), where the thermal ground state is a mixture of entangled states (e.g., the singlets) and product states (e.g., the fully polarized state)~\cite{Vidal2004Entanglement_first}. Thirdly, whenever the lower bound becomes zero, we can often find a separable state close to the thermal state at the corresponding temperature. This shows that we can characterize the entanglement threshold temperature very accurately with the spin-squeezing parameter, even for small $N$, which should not necessarily be expected from the asymptotic completeness argument. Furthermore, we observe that in some cases when the ground state is separable, entanglement arises at higher temperatures. In our case, we witness this in the FM XX model 
(cf. \cref{fig:LMG_FM}). Here, we must also note that in the re-entrant region the upper bound is often extremely loose and its exact value is not very meaningful (close to 1 for $N = 8$). Hence, our upper bound works best for detecting full separability or large entanglement, while it often considerably overestimates weak entanglement.

\begin{figure}
    \centering
    \includegraphics[width=.6\textwidth]{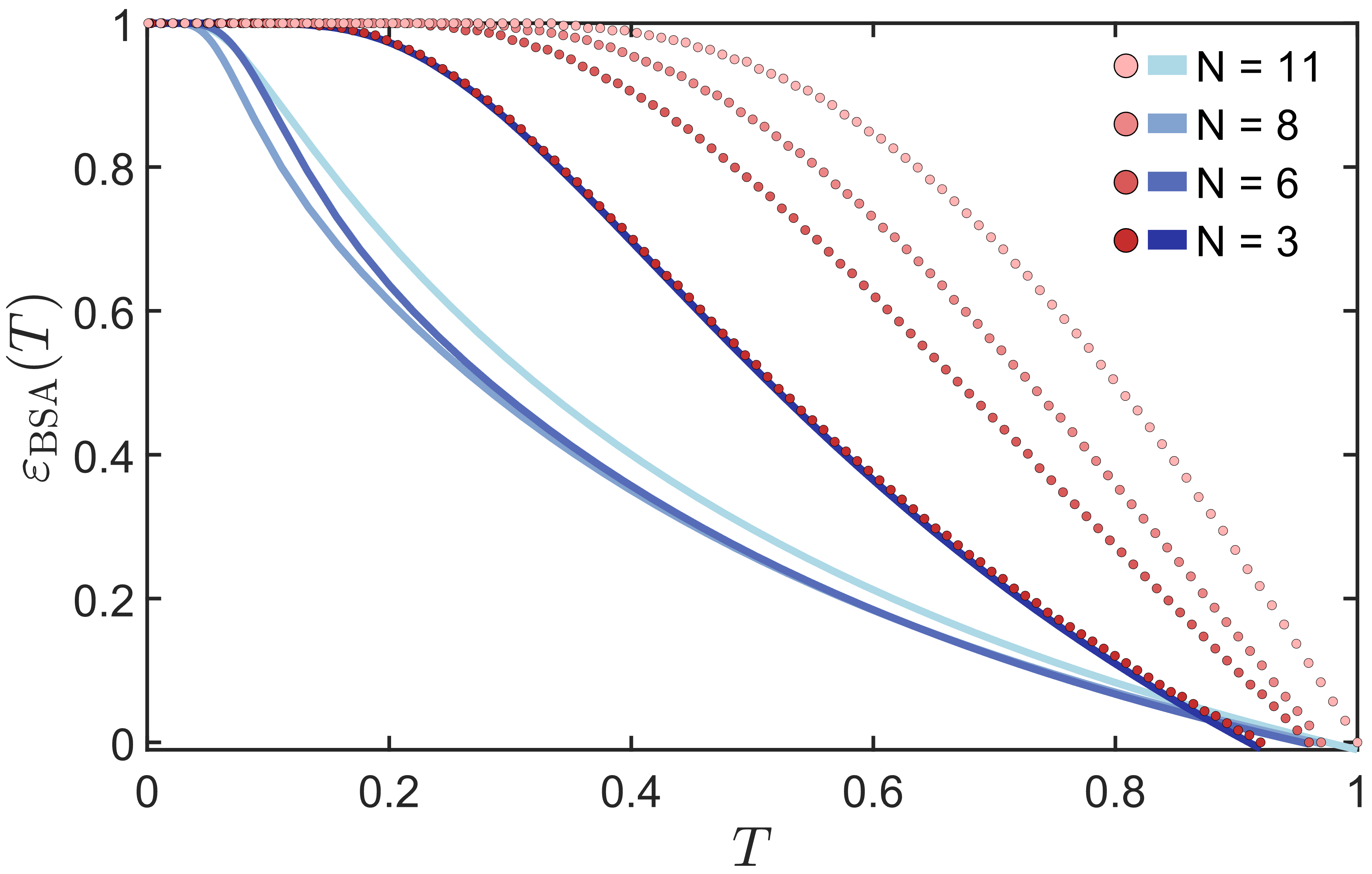}
    \caption{Lower (blue) and upper (red) bounds for the BSA of thermal states of the XXX model.
    The lower bounds depend monotonically on $N$ for $N$ either odd or even but not in general.
    }
    \label{fig:BSA_diff_N}
\end{figure}

Considering the upper bounds more closely, we find that in the case of $N=3$ qubits in the XXX model we can see that the upper and lower bound to the BSA coincide {\it at all temperatures} (cf. \cref{fig:BSA_diff_N}), which is quite surprising, given that already for three qubits finding the exact value of the BSA is generally a very hard problem ~\cite{Akulin_2015}. Note that this also happens if one considers a different entanglement monotone, i.e., the generalized robustness as illustrated in \cref{app:GR}. However, in general the bounds are non-tight for generic entangled thermal states  (cf. \cref{fig:BSA_diff_N}, \cref{fig:LMG_lower_bounds}) for $N > 3$. To some extent, this is simply due to our limited computational resources, and the difficulty in searching for the closest separable state, as well as to the fact that the lower bound might be itself non-tight in those cases.

More specifically, we believe the mismatch is mostly due to the overlap computation of the upper bound algorithm that is based on maximizing the overlap with single-particle reduced density matrices. For $N = 3$, this is still a good way to sample the full state space; for larger $N$ we suspect that one would need to also include optimization based on $2, 3, ...$-party marginals which is very difficult to implement efficiently. Another insight concerning the upper bound is that it does not reach zero in the AFM case for $h > 0$, as in \cref{fig:LMG_AFM}. The reason for this is twofold: One, the relevant states in the AFM case are no longer (mostly) constrained to the symmetric subspaces (where they have a very simple structure) but have more significant contributions also from the anti-symmetric and mixed subspaces - this makes it much harder to find tight upper bounds. Secondly, it might be the case that there is some entanglement that is simply not captured by the SSI witness in this case, meaning that full separability is not yet reached. Regarding \cref{fig:LMG_AFM}, we also observe that in some cases the upper bounds computed from the simplified ansatz in \eqref{eq:simpleans} perform better than those computed using the full ansatz in \eqref{eq:fullansatz}, which shows that in a few cases a very simple ansatz captures the optimal separable state quite well and converges more rapidly to a better approximation than the full ansatz that can get stuck more easily.

\begin{figure}[htp]
\begin{tcolorbox}[colback=white,colframe=LightBlue!,title=\textbf{\large Understanding the XXZ model: Phases and Spin-Squeezing}]
 The entanglement detected by the SSIs resembles the ground state phase diagram but shows that for a given phase, entanglement is not uniform; i.e. closer to the fully polarized phase, states are almost polarized and entanglement is weaker. Observing that the singlet-dominated AFM region and the Dicke-like region have very similar values of the SSI parameter, we find that different phases can host similar entanglement. For larger $T$, the entanglement smears out due to thermal fluctuations. 
 
    \begin{subfigure}[b]{.49\linewidth}
        \includegraphics[width=1\linewidth]{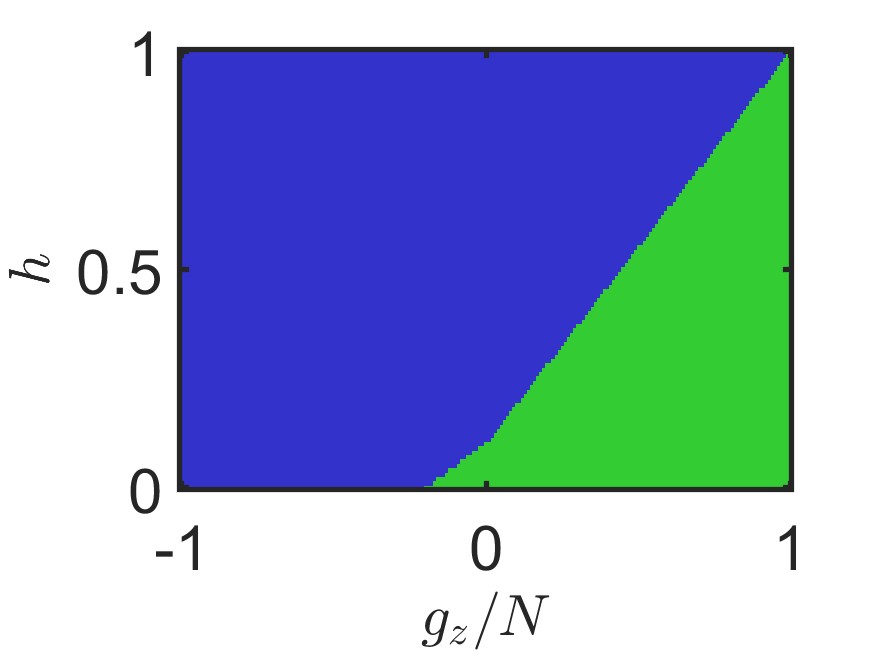}
    \end{subfigure}
    \qquad
    \begin{subfigure}[b]{.49\linewidth}
        \includegraphics[width=\linewidth]{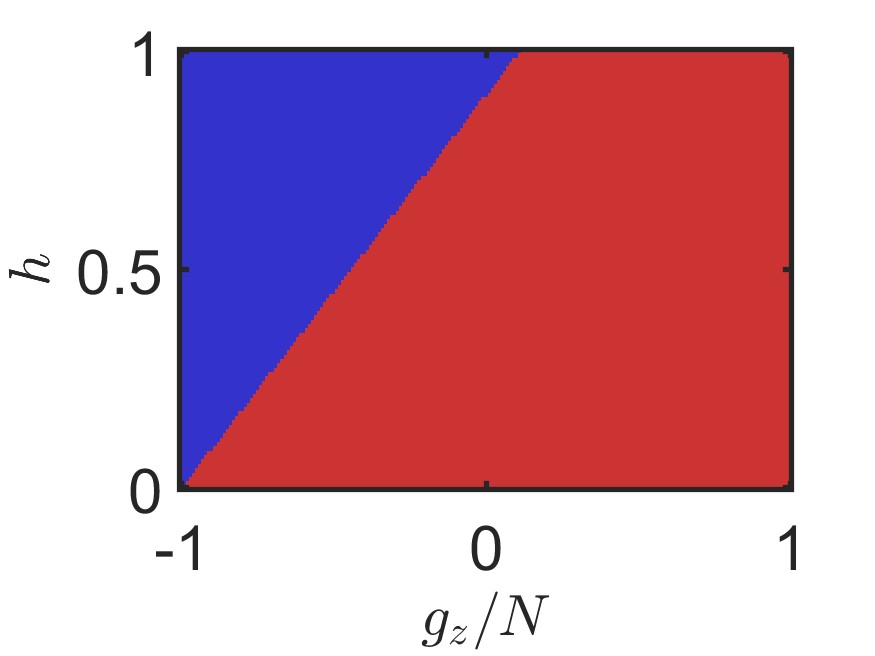}
    \end{subfigure}
    \begin{subfigure}[b]{.49\linewidth}
        \includegraphics[width=1\linewidth]{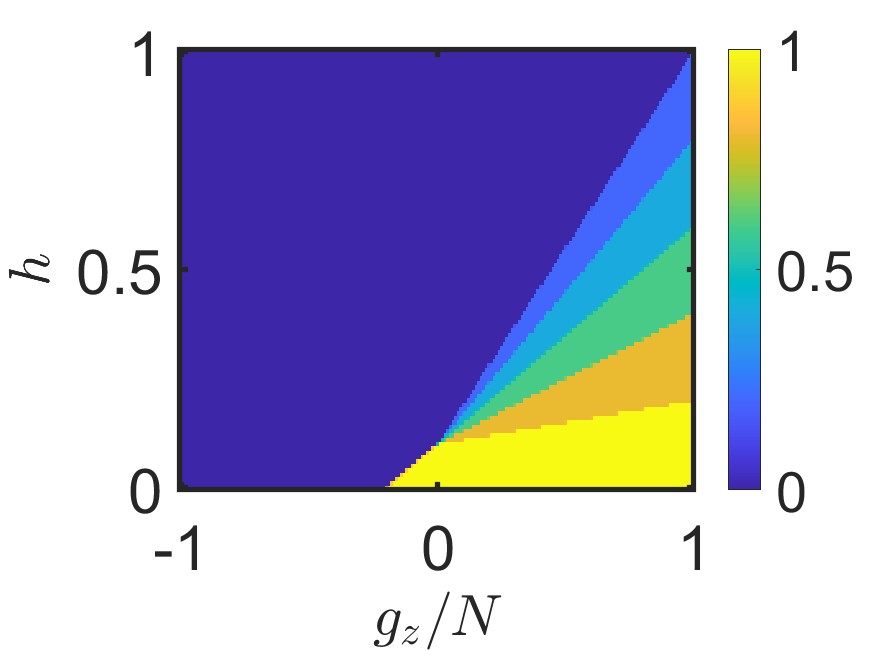}
    \end{subfigure}
    \qquad
    \begin{subfigure}[b]{.49\linewidth}
        \includegraphics[width=1\linewidth]{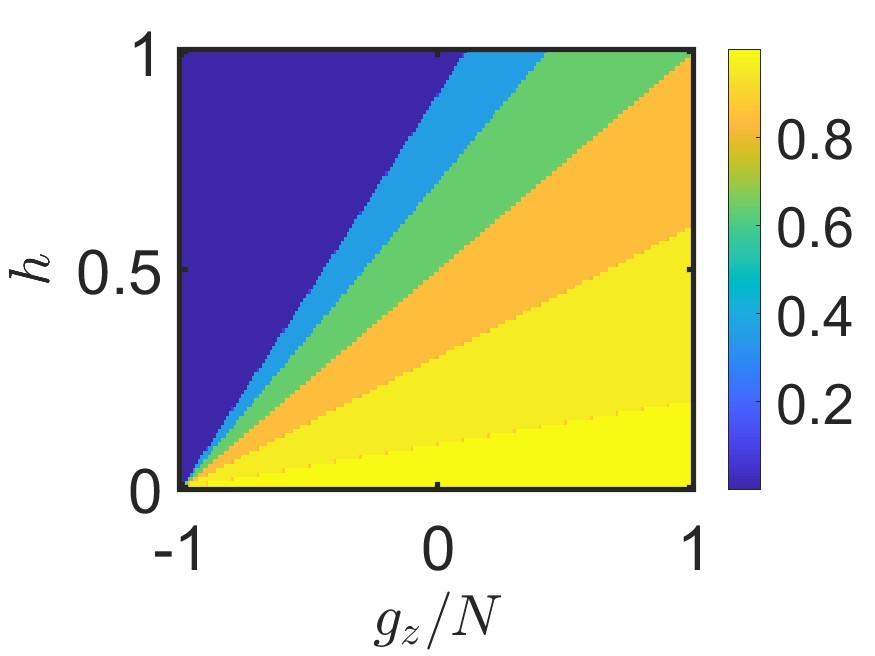}
    \end{subfigure}
    \begin{subfigure}[b]{.49\linewidth}
        \includegraphics[width=1\linewidth]{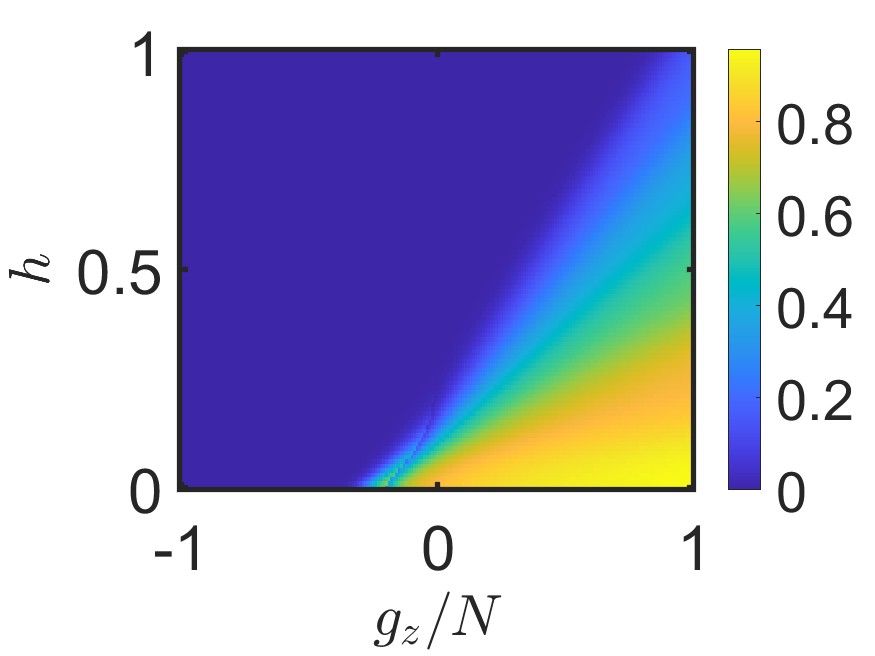}
    \end{subfigure}
    \qquad
    \begin{subfigure}[b]{.49\linewidth}
        \includegraphics[width=1\linewidth]{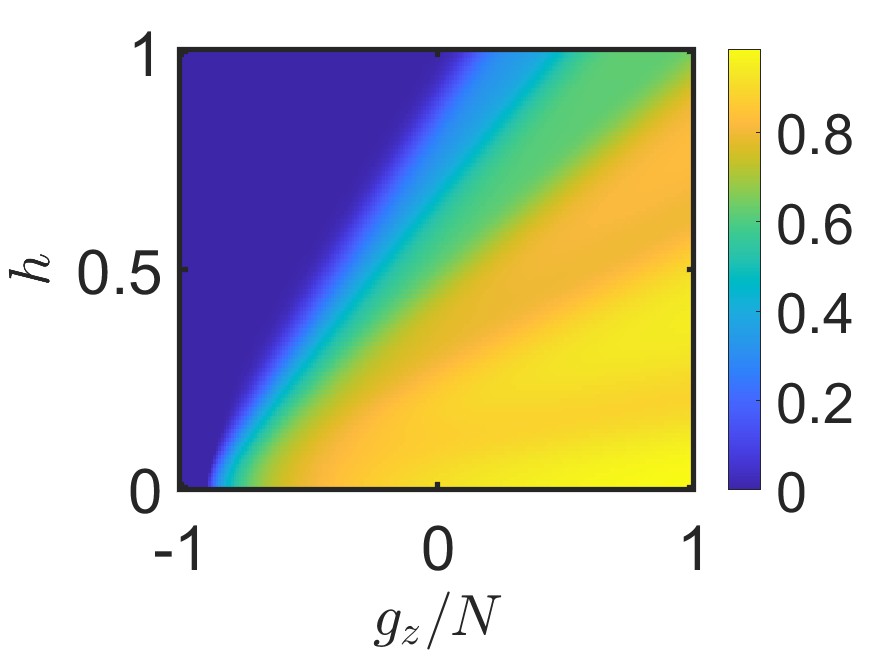}
    \end{subfigure}
    \caption{Top row: Ground-state phase diagram of the AFM (left) and FM (right) XXZ model; we classify the following ground-state phases: fully polarized (blue), singlet states (green), and unpolarized Dicke state (red). Center row: Entanglement as quantified by the SSI parameter (normalized as a lower bound to the BSA) for a thermal state at $T = 0.001$ (close to ground state) for the model above. Bottom row: Entanglement as quantified by the SSI parameter for a thermal state at $T = 0.5$ (intermediate-$T$ thermal state).}
    \label{fig:phases_SSI}
\end{tcolorbox}
\end{figure}

Finally, as an overview and to illustrate the usefulness of the SSI-based lower bounds to the BSA, consider \cref{fig:phases_SSI} where we investigate how entanglement evolves across different phases in the FM and AFM XXZ model. While the phase diagram\footnote{Note that in our model, the finite-$N$ diagram already looks like the thermodynamic one, which is why we call our numerically obtained plots phase diagrams.} is calculated for the ground state at $T = 0$, we can create similar plots based on the SSI witnesses. At each value of $T$, the SSI diagrams are much richer and tell us something about the states and differently entangled regions within a given phase. 

\subsection{Analytic lower bounds for special cases}
\label{sec:analytic}

As we have seen, the spin-squeezing parameter can be computed numerically easily for relatively large $N$ (cf. \cref{fig:LMG_lower_bounds}). In some cases it is even possible to find an analytic scaling expression valid for large $N$, expressing the partition function in an integral form.
The simplest model to consider is the fully connected Heisenberg model with equal couplings, i. e., $g=g_Z=1$ and $h=0$, which was also considered previously in \cite{tothpra05}.
In the thermodynamic limit of $N \rightarrow \infty$, we can express the partition function in \cref{eq:partition_function} as an integral:
\be
    \mathcal Z_T \simeq \int_{0}^{1} \de x \ e^{-\frac{N}{2} x(x+1)} (1 + Nx)^2 \frac{2^n \left(\frac{1}{N}\right)^{5/2} (4 N-9)}{\sqrt{2\pi }} \mathrm{e}^{- g \frac{N}{4} x(x+1)/T} ,
\ee
which can be integrated with symbolic numerical programs. 
Here we have defined the real variable $x:= 2J/N$ 
and used the multiplicity expansion 
\be\label{eq:muxexpansion}
\mu_J \rightarrow \mu_x \approx  e^{-\frac{N}{2} x(x+1)} (1 + Nx) \frac{2^n \left(\frac{1}{N}\right)^{5/2} (4 N-9)}{\sqrt{2\pi }} ,
\ee
derived in \cite{Curtright_2017}.
In this case, the spin squeezing parameter in \cref{eq:SSIspar} can be easily calculated as we have $\aver{J_x} = \aver{J_y} = \aver{J_z} = 0$ and $\tr(\Gamma_\varrho) = \aver{H}_\varrho$. Moreover, due to full rotation invariance we also have $(\mathfrak X_\varrho)_{xx} = (\mathfrak X_\varrho)_{yy} = (\mathfrak X_\varrho)_{zz}$ and thus they are either all positive or all negative. In the latter case, the spin-squeezing parameter is strictly positive (and thus no entanglement is detected), while in the former case we have 
\be
\xi_{SS}(T) =(\Delta J_x)_{\varrho_T}^2 + (\Delta J_y)_{\varrho_T}^2 + (\Delta J_z)_{\varrho_T}^2 - \frac N 2 = \aver{H}_T -  \frac N 2 .
\label{eq: SSI_witness_energy}
\ee
In turn, the mean energy can be readily calculated from the partition function as
\be
    \aver{H}_T  = - N \partial_\beta \log{\mathcal Z_T} , 
\ee
and from the integral expression for the partition function one can even calculate it in the thermodynamic limit $N \rightarrow \infty$. 
We obtain
\be
\aver{H}_T = \frac{3NT}{4T + 2} \implies \xi^{N \gg 1}_{SS}(T) = \frac{3NT}{4T + 2} - \frac{N}{2},
\ee
and immediately see that the lower bound to the BSA vanishes for $T = 1$ (in general $T=g$ for an arbitrary coupling scale). This result coincides with what is found numerically, as described in \cref{sec:numerical_results}. See also \cite{tothpra05} for a similar scaling calculation and \cite{tothPRA09} for a similar numerical temperature bound for small values of $N$.
We can then write our lower bound to the BSA asymptotically as (setting $g = 1$ for simplicity)
\be
\mathcal E^{N \gg 1}_{BSA}(T) \geq - \frac 2 N \xi^{N \gg 1}_{SS}(T) = \max \left\{0, 1 - \frac{6T}{4T + 2} \right\} ,
\ee
which is remarkably independent of $N$ (at lowest order).

Let us now consider a less symmetric and more rich model, that is the XX model with an external field.
In this case, the symmetry of rotations is reduced to those about the $J_z$ axis.
The partition function can be once again approximated with an integral expression in the limit $N\rightarrow \infty$:
\be
\mathcal Z_T \simeq \int_{0}^{1} \de x \int_{-x}^{x} \de y \ \mu_x \ \mathrm{e}^{- \frac{gN}{4} \left(x(x+1) - y^2 - 2\tfrac h g y\right)/T} ,
\ee
where we now introduced a new variable $y:= 2J_z/N \in [-x, x]$, which leads to a double-nested Gaussian integral, and used the same multiplicity expansion as in \cref{eq:muxexpansion}.

We consider the AFM case $g=1$, that has some analogies with the XXX model discussed previously.
In this case, there is a first-order QPT in the limit $N\rightarrow \infty$ at zero external field and we can analytically calculate the lower bound to the BSA in the case $g=1$ and $h=0$. Similarly as for the XXX case, this is given by
\be
\mathcal E^{g=1,h=0}_{BSA}(T) \geq  \max \left\{0, 1 - \frac 2 N \left[(\Delta J_x)_{\varrho_T}^2 + (\Delta J_y)_{\varrho_T}^2 + (\Delta J_z)_{\varrho_T}^2\right] \right\} ,
\ee
where however the trace of the covariance matrix $\tr(\Gamma_{\varrho_T}) = (\Delta J_x)_{\varrho_T}^2 + (\Delta J_y)_{\varrho_T}^2 + (\Delta J_z)_{\varrho_T}^2$ is not anymore simply given by the energy.
In this case, the large $N$ approximation results in
\be
(\Delta J_x)_{\varrho_T}^2 + (\Delta J_y)_{\varrho_T}^2 + (\Delta J_z)_{\varrho_T}^2 \simeq \frac{2 NT}{2T+1} ,
\ee
thus implying that the lower bound to the BSA is
\be
\mathcal E^{g=1,h=0}_{BSA}(T) \geq \max \left\{0, 1 - \frac{4 T}{2T+1} \right\},
\ee
and vanishes for $T = 1/2$. Once again, this result shows a constant scaling with $N$ (at lowest order) and coincides with what we found numerically in \cref{sec:numerical_results} for finite values of $N$ (and what was found in \cite{tothPRA09} for small $N$).

\section{Conclusion and Outlook}\label{sec:conclusions}

In conclusion, we have investigated how to quantify entanglement in mixed many-body states, with a particular focus on permutationally and rotationally invariant states that 
arise in the context of spin-squeezing. We have investigated how to find lower and upper bounds to the best separable approximation, which quantifies the distance from the set of fully separable states, and thus in some sense the deviation from a mean-field ansatz. We applied our methods to thermal states of fully-connected spin models by using a well-known set of SSIs (which are nonlinear in the quantum state). 

While the lower bounds obtained from SSIs are directly computable from collective observables and scale very favorably with system size, the corresponding upper bounds rely on explicit state constructions and are not guaranteed to be tight in all regimes. Improving these upper bounds remains an interesting direction for future work. In contrast, the lower bounds already provide a robust and scalable entanglement quantifier that can be directly connected to experimentally measurable quantities.
Results of this paradigmatic application complements existing literature on entanglement of ground states, and in particular extends those to the case of non-zero temperature states. We thus showed how to {\it quantitatively} distinguish quantum from classical correlations in situations where the former become important
for describing the physics of the system. For example, we observe that even when the ground-state is separable, the first excited states can be highly entangled, which
is then reflected into entanglement of the thermal state, at least at relatively small temperatures.
This argument suggests an investigation of the behavior of entanglement in thermal states also in other situations, such as excited state quantum phase transitions (ESQPTs)~\cite{Caprio2008Excited,Vidal2004Entanglement_second,Vidal2004Entanglement_first} or usual thermal phase transitions.

Importantly, the lower bound derived in this work relies only on collective second moments, which can be related to static response functions via fluctuation-dissipation relations \cite{HaukeHeylTagliacozzoZoller16}. In particular, collective variances and covariances can be obtained from static susceptibilities to global fields, independently of whether the underlying Hamiltonian is permutation-invariant or quadratic in the collective spin operators. Therefore, our lower-bound construction can in principle be applied to a broader class of many-body systems than just the models considered here. However, in systems without collective symmetry, other witnesses tailored to local structure may outperform the purely collective SSIs.

In this sense, our work unlocks further potential investigation related to entanglement and many-body physics, including on the relation between thermodynamic quantities, entanglement measures
and quantum phases in models with different symmetries. 
One follow-up question then becomes how to efficiently look for optimal sets of such thermodynamic entanglement witnesses. This could lead to an improved classification of phases of matter, that takes into account entanglement (i.e., quantum correlations) also beyond the ground state.
This question becomes particularly relevant in out-of-equilibrium situations, where entanglement represents precisely the bottleneck for a deeper investigation and classification of the various phases, while its role associated to physically relevant quantities remains still elusive~\cite{KirtonRosesKeelingDallaTorre2018DickeReview,DEFENU20241}.

\begin{figure}[htp]

\begin{tcolorbox}[colback=white,colframe=DarkBlue!,title=\textbf{\large Outlook: Global interaction quenches}]

One-axis or two-axis squeezing \cite{KitagawaUeda1993, SoerensenNAT2001}: At time $t = 0$, a collective interaction is suddenly switched on, driving the system out of equilibrium from an initially separable state. For finite system sizes, the entanglement exhibits oscillatory behavior and revivals. Note that $\aver{W}_{\rm SSI} = \xi_{SS}$ is the rescaled optimal SSI witness that is the lower bound to the BSA.

\centering 
\begin{subfigure}[b]{.45\linewidth}
    \includegraphics[width=1\linewidth] {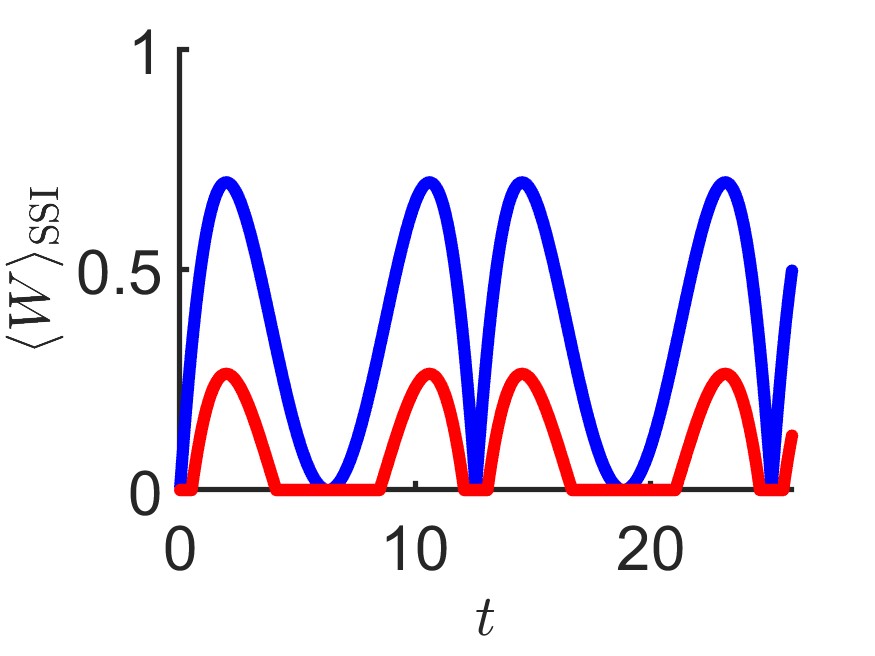}
\end{subfigure}
\qquad
\begin{subfigure}[b]{.45\linewidth}
    \includegraphics[width=\linewidth]{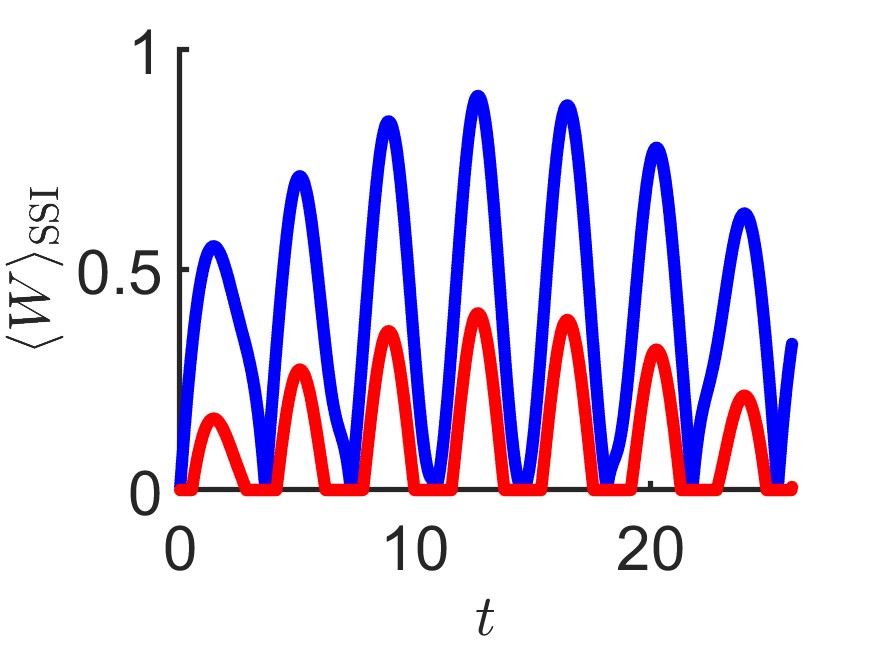}
\end{subfigure}
\begin{subfigure}[b]{.45\linewidth}
    \includegraphics[width=\linewidth]{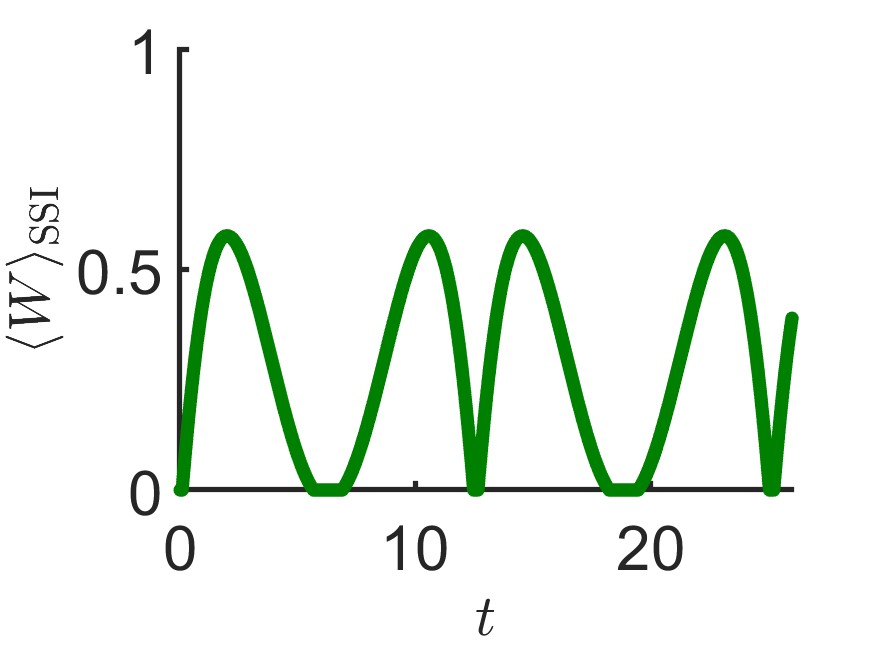}
\end{subfigure}
\qquad
\begin{subfigure}[b]{.45\linewidth}
    \includegraphics[width=\linewidth]{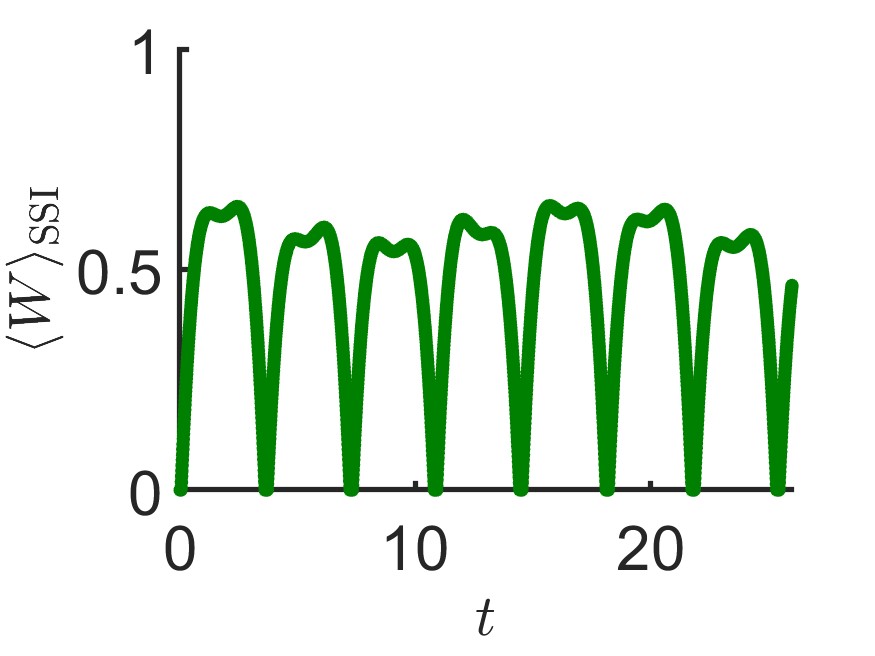}
\end{subfigure}
\caption{Left: One-axis-squeezing with $H_{\rm OAT} = \frac{g_z}{N} J_z^2$. Right: Two-axis-squeezing with $H_{\rm TAT} = \frac{g}{N} (J_x^2-J_y^2)$. Top row: Results for the fully polarized state along $J_x$, without noise (blue), and with noise (red): $\rho_{\rm noisy} = (1-\varepsilon) \ketbra{\Psi_{\rm pol}} + \varepsilon \mathbb{1}/d$, $\varepsilon = 0.3$. Bottom row: Thermal state with $g = 1, g_z = -1, h = 0.5$ at $T = 0.3$.}
\label{fig: outlook}
\end{tcolorbox}
\end{figure}

Referring to the particular case treated in this work, while the relationship between spin-squeezing and entanglement is well established in equilibrium and near-equilibrium settings, a systematic understanding of SSIs as entanglement quantifiers in far-from-equilibrium collective dynamics is still missing. As an example of such a system, in \cref{fig: outlook} we analyze what happens in the case of global dynamical quenches, starting from a mixed initial state, either at equilibrium or with generic noise, and tracking the evolution of entanglement in out-of-equilibrium states. This shows how we can use the results derived in this work to investigate the evolution of entanglement also in out-of-equilibrium scenarios that are close to experimental settings, which can potentially help in understanding and classifying complex non-equilibrium phase transitions also in relation to experiments~\cite{lewis2019unifying,li2023improving,mann2025squeezingclassicalantiferromagnetsquantum}.

As a final comment, we emphasize the fact that it is clear that symmetries play a crucial role, and that it is thus important to investigate physical systems with different degrees of symmetry. For example, relaxing the permutational symmetry of the systems considered here to just translational symmetry, and once more complementing existing literature on the scaling of entanglement in ground states or out-of-equilibrium pure states~\cite{amico08,Laflorencie16}.
It is obvious that in less symmetric systems it is generally much harder to calculate any entanglement measure. Thus, the question arises on how to further improve our methods to tackle the most general case of no symmetries at all, at least for systems of the order of $N \approx 10-12$ qubits, where results with relevance for many-body systems can already be obtained. 

\section*{Acknowledgments}
This research was funded in whole or in part by the Austrian Science Fund (FWF) [10.55776/P35810]. For open access purposes, the author has applied a CC BY public copyright license to any author accepted manuscript version arising from this submission. GV and JM also acknowledge support from the grant P 36633-N (Stand-Alone). GV further acknowledges support from the grant No. RYC2024-048278-I funded by MCIU/AEI/10.13039/501100011033 and FSE+.
AU is financially supported by JSPS Overseas Research Fellowships and acknowledges financial support from Spanish MICIN (projects: PID2022:141283NBI00; 139099NBI00) with the support of FEDER funds, the Spanish Government with funding from European Union NextGenerationEU (PRTR-C17.I1) and the Generalitat de Catalunya.
OG acknowledges support by the  Deutsche Forschungsgemeinschaft (DFG, German Research Foundation, project numbers 447948357 and 440958198),
the Sino-German Center for Research Promotion (Project M-0294), the German Ministry of Education and Research (Projects QuKuK, BMBF Grant No.~16KIS1618K) and EIN Quantum NRW.

\appendix
\begin{widetext}

\section{Permutationally and rotationally invariant separable states}\label{app:separablePIstates}

Here, we discuss in more detail how to describe separable states that have invariance under {\it particle permutations} and {\it global rotations}. 
First of all, let us recall that separable states are those that can be decomposed as
\be
\varrho_{\rm sep} = \sum_k p_k (\ketbra{\psi_1} \otimes \dots \otimes \ketbra{\psi_N})_k ,
\ee
where $\ketbra{\psi_j}$ are pure single-particle states and $p_k$ is a probability distribution. Thus, separable states are just the convex hull of pure product states
\be\label{eq:prodstates}
\ketbra{\Psi_{\rm prod}} = \ketbra{\psi_1} \otimes \dots \otimes \ketbra{\psi_N} .
\ee
Now, a way to write states that are separable and also permutationally invariant (PI) would be to first consider the boundary of the set of separable states, i.e. product states, and map those into PI states, which will be mixed.
Those (mixed) PI states would be the boundary of the set of separable PI states. Any other separable PI state can be then written as a convex combination of boundary states~\cite{VollbrechtWerner2001}.

Permutationally invariant states $\varrho$ are such that
\be
U_\pi \varrho U^\dagger_\pi = \varrho ,
\ee
where $U_\pi$ is any unitary operator associated to a permutation of the particles $\pi \in \mathfrak{S}_N$ ($\mathfrak{S}_N$ being the permutation group of $N$ elements).
Concretely, the action of a unitary permutation is given by
\be
U_\pi \ket{\psi_1} \ket{\psi_2} \dots \ket{\psi_N} = \ket{\psi_{\pi(1)}}\ket{\psi_{\pi(2)}}\dots\ket{\psi_{\pi(N)}}.
\ee

Then the boundary states of the set of PI separable states are constructed by taking a generic pure product state as in \cref{eq:prodstates}
and mixing it with all its permuted versions, i.e.:
\be\label{eq:prodPIdefi}
\varrho_{\rm prod, PI} = \frac 1 {N!} \sum_{\pi \in \mathfrak{S}_N} U_{\pi} \ketbra{\psi_1} \otimes \dots \otimes \ketbra{\psi_N} U_{\pi}^\dagger ,
\ee
which is a generic state belonging to the boundary of PI separable states (which is why we labeled it as ``product'', even if it is not a pure product state).
Note also that two product states that are a permuted version of each other, i.e. $\ket{\Psi^\prime_{\rm prod}}= U_\pi \ket{\Psi_{\rm prod}}$ are mapped into the same separable PI state.

To express such states, we do not need to actually mix the state with all its permuted versions, but we can exploit the fact that every PI state can be expressed as
a direct sum of blocks pertaining to different collective spin sectors, due the well-known Schur-Weyl duality. In short, for our purposes this result states that under the action of operators of the form $A^{\otimes N}$, the $N$-qubit Hilbert space can be partitioned as a direct sum
\be\label{eq:hilspacDec}
\mathcal H_N = \bigoplus_{J=0}^{N/2} \mathbb C^{2J+1} \otimes \mathbb C^{\mu_J} ,
\ee
where $\mathbb C^{2J+1}:=\mathcal H_J$ is a Hilbert space of a single spin-$J$ particle with $0\leq J\leq \tfrac N 2$ and $\mu_J =\binom{N}{N/2-J}-\binom{N}{N/2-J-1}$ is the corresponding multiplicity.
Each sector $\mathcal H_{J}$ contains states that belong to a representation
of the symmetric group given by an associated Young tableau with $N$ boxes and at most two rows.
For example, the sector with $J=\tfrac N 2$, which has dimension $D=2J+1=N+1$ is composed by states that are mapped onto themselves (with factor $+1$) by all possible permutations $U_\pi$. 
This is generated by the so-called Dicke states and corresponds to the Young diagram with one single row and $N$ columns, and has always multiplicity $\mu_{N/2}=1$.
Instead, the sector $J=0$ with dimension $D=1$ is composed by a singlet state, which is a product of two-body singlets corresponding to a Young diagram with two rows and $N/2$ columns, e.g., $\ket{J=0,i_0=1} = \ket{\Psi_{12}^-}\ket{\Psi_{34}^-} \dots \ket{\Psi_{N-1 N}^-}$ with
\be
\ket{\Psi_-} = \tfrac 1 {\sqrt{2}} \left( \ket{\hspace{-3pt}\uparrow \downarrow} - \ket{\hspace{-3pt}\downarrow \uparrow} \right) ,
\ee
being the two-qubit singlet state. However, there are many possible such $N$-qubit singlets, corresponding to the different possible Young tableaux with two rows and $N/2$ columns, which
gives the multiplicity: $\mu_0 =\binom{N}{N/2}-\binom{N}{N/2-1}$. In other words, the multiplicity of the $J=0$ sector corresponds to the number of possible ways of making tensor products of $2$-body singlets.

Now, let us consider collective operators. For example, given the single-qubit spin operators $j_k = \tfrac 1 2 \sigma_k$ with $k \in \{x,y,z\}$ we construct the collective spin components as
\be\label{eq:collSpinopDef}
J_k = \sum_{n=1}^N j_k^{(n)} ,
\ee
and they split into sectors as provided by the Hilbert space decomposition in \cref{eq:hilspacDec}, each sector $\mathcal H_J$ corresponding to a different irreducible representation of $su(2)$. In particular, the Casimir operator given by the squared total spin is proportional to the identity in each sector, namely
\be\label{eq:totalSpin}
J_x^2 + J_y^2 + J_z^2 = \bigoplus_{J=0}^{N/2} J(J+1)\id_{J} \otimes \id_{\mu_J} ,
\ee
where $\id_{J}$ and $\id_{\mu_J}$ are identity operators in $\mathcal H_J$ and $\mathbb C^{\mu_J}$ respectively.

Because of the decomposition \eqref{eq:hilspacDec}, density matrices that are permutationally invariant can be decomposed as 
\be\label{eq:PIstatesgeneral}
\varrho = \bigoplus_{J=0}^{N/2} p_J \varrho_J \otimes \tfrac 1 {\mu_J} \id_{\mu_J} ,
\ee 
where each $\varrho_J$ belongs to a different spin sector and $\{p_J\}$ is a probability distribution, i.e. $p_J \geq 0$ and $\sum_J p_J=1$.

Because of this decomposition, all we need to describe any PI state is to find its coefficients in each of the spin-$J$ sectors. 
Essentially, this is because PI states do not carry any coherence between different spin-$J$ subspaces. Moreover, they also do not carry any coherence between the various subspaces with the same $J$ and different indices in the multiplicity space $1\leq i_J \leq \mu_J$. In addition, their matrix elements in the given spin-$J$ subspaces are independent of $i_J$. All of this can also be seen from the general decomposition in \cref{eq:PIstatesgeneral}.
Thus, in summary we are interested in the expectation values of the operators 
\be 
\sum_{i_J=1}^{\mu_J}\kb{J , J_z , i_J}{J , J^\prime_z , i_J} = \kb{J , J_z}{J , J^\prime_z} \otimes \id_{\mu_J} , 
\ee 
where $\ket{J , J_z , i_J}$ is a pure state that is the common eigenstate of $J_x^2 + J_y^2 +J_z^2$ and $J_z$, and $i_J$ is the index in the multiplicity space that labels the different states with the same value of $J$ and $J_z$ but a different Young tableau corresponding to the given $J$. 

Then, considering a product state $\ket{\psi_1 \dots \psi_N}$ we can find its PI twirled version $\varrho_{\rm prod, PI}$ as in \cref{eq:prodPIdefi} from the following coefficients

\begin{align}
\alpha^{\rm prod}_{J , J_z , J^\prime_z} & := \tr\left( \varrho_{\rm prod, PI} \sum_{i_J} \kb{J , J_z , i_J}{J , J^\prime_z , i_J} \right) \\
& = \frac 1 {N!} \sum_{\pi \in \mathfrak{S}_N} \tr\left(  \ketbra{\psi_1} \otimes \dots \otimes \ketbra{\psi_N} U_{\pi}^\dagger \kb{J , J_z}{J , J^\prime_z} \otimes \id_{\mu_J} U_{\pi} \right) \\ 
& = \tr\left(  \ketbra{\psi_1} \otimes \dots \otimes \ketbra{\psi_N} \kb{J , J_z}{J , J^\prime_z} \otimes \id_{\mu_J} \right),
\end{align}

where we dropped the index corresponding to $i_J$. As mentioned, that is because for PI states we need to consider only the operators $\kb{J , J_z}{J , J^\prime_z} \otimes \id_{\mu_J}$, which are already PI.

A further simplification arises for a product state that is a product of eigenstates of the single particle $j_z$ observable. In that case, the coefficients above become
\be\label{eq:SSzSzprimeofmk}
\alpha_{J , J_z , J^\prime_z}^{m_1,\dots , m_N} =  \sum_{i_J} \langle m_1, \dots , m_N \kb{J , J_z , i_J}{J , J^\prime_z , i_J} m_1, \dots , m_N \rangle .
\ee
Moreover, due to permutation invariance two states that are a permuted version of
each other have the same coefficients, i.e., 
\be
\alpha_{J , J_z , J^\prime_z}^{m_1,\dots , m_N} = \alpha_{J , J_z , J^\prime_z}^{\pi(m_1,\dots , m_N)} ,
\ee
where $\pi(m_1,\dots , m_N)$ is any permutation of the sequence $(m_1,\dots , m_N)$. 
Thus, in this case we can just put a label for the number $K$ of $+1$ eigenvalues, because, once more, two states $\ket{m_1,\dots , m_N}$ and $\ket{m^\prime_1,\dots , m^\prime_N}$
with the same number of values $m_k=1$ are mapped into the same PI separable state.
Moreover, a product spin state with $K$ spins pointing up in the $j_z$ direction is an eigenstate of $J_z$ with eigenvalue $M_z=2K - N$. 
In fact, each of the states $\ket{J , M_z, i_J}$ is a superposition of the corresponding states $\ket{\pi(m_1,\dots , m_N)}$.
Because of that, we have the following relation (valid for all $J \geq |M_z|$)
\begin{align}
\label{eq:alphaSSzpim1N}
\sum_{\pi} \alpha_{J , J_z , J^\prime_z}^{\pi(m_1,\dots , m_N)} & = \sum_{\pi} \sum_{i_J} \langle \pi( m_1, \dots , m_N ) \kb{J , J_z , i_J}{J , J^\prime_z , i_J} \pi( m_1, \dots , m_N ) \rangle \\
& = \delta_{J_z M_z} \delta_{J^\prime_z M_z}  c_{M_z} = \delta_{J_z M_z} \delta_{J^\prime_z M_z}  \frac{N!}{I_{M_z}},
\end{align}
where the factor $c_{M_z}$ is simply calculated from combinatorics, and given by the total number of permutations $N!$ divided by the number of distinct states with the same value of $J_z=M_z$, which we termed $I_{M_z}$. 
Thus, since all coefficients $\alpha_{J , J_z , J^\prime_z}^{\pi(m_1,\dots , m_N)}$ are equal to each other, we simply obtain that each of them is equal to
\be\label{eq:alphaJJzapp}
\alpha_{J , J_z , J^\prime_z}^{m_1,\dots , m_N} =  \left\{ 
\begin{array}{c}
  \delta_{J_z M_z} \delta_{J^\prime_z M_z}  \frac{1}{I_{M_z}} \quad \text{for} \ J \geq |M_z| ,    \\
     0 \quad \text{for} \ J \leq |M_z| ,
\end{array}
\right.
\ee
which is obtained from \cref{eq:alphaSSzpim1N} by dividing with the total number of permutations.

Let us now consider states that are invariant under permutations and global rotations around the $J_z$ axis.
By definition we then have that
\be
\mathcal P_z(\varrho_{\text{PI},z}) = \varrho \quad \text{with} \quad \mathcal P_z(\varrho_{\text{PI},z}) = \int_{0}^{2\pi} \de \theta \ e^{-i\theta J_z} \varrho_{\text{PI},z} e^{i\theta J_z} . 
\ee
As a consequence, such a state can be always expanded as 
\be
\varrho_{\text{PI},z} = \sum_{J, J_z} \alpha_{J , J_z} \ketbra{J,J_z} \otimes \id_{\mu_J} ,
\ee
where the expansion coefficients are simply
\be\label{eq:alphaSSzprod}
\alpha_{J, J_z}^{\rm prod} = \sum_{i_J=1}^{\mu_J} | \langle \phi_1, \dots , \phi_N \ket{J,J_z,i_J}|^2 ,
\ee
which are obtained from the overlaps with the common eigenstates of $J_x^2 + J_y^2 +J_z^2$ and $J_z$.

The coefficients in \cref{eq:alphaSSzprod} can be further expanded by inserting the resolution of identity in a reference ``computational'' basis, like the product of local $j_z$ eigenbases. Then we get
\begin{align}
    \alpha_{J, J_z}^{\phi_1, \dots , \phi_N} & = \sum_{i_J=1}^{\mu_J} | \langle \phi_1, \dots , \phi_N \ket{J,J_z,i_J}|^2 \\
    & = \sum_{i_J=1}^{\mu_J} \left| \sum_{m_1,\dots , m_N}  \left\langle \phi_1, \dots , \phi_N \ketbra{m_1 \dots m_N}  J ,J_z,i_J \right\rangle \right|^2 \\
    & = \sum_{i_J=1}^{\mu_J} \left| U(\phi_1 , \dots , \phi_N) U_{\rm Schur}^\dagger  \right|^2 ,
\end{align}

where on the right-hand side we get a matrix product between an $N$-particle local rotation and the so-called {\it Schur matrix},
defined as
\be
U_{\rm Schur} := \kb{J,J_z,i_J}{m_1,\dots , m_N} .
\ee
Note however that the Schur matrix becomes computationally quite hard to calculate for larger and larger $N$~\cite{Kirby_2018}.

\section{Bounding the BSA from spin-squeezing inequalities}\label{app:boundfromSSIs}

Here let us explain in more detail how to calculate lower bounds to the best separable approximation from the SSIs in \cref{eq:GenSSIsFull}. 
Let us first recall how to get a lower bound to the BSA (see also \cite{FadelVitagliano_2021}). From the dual definition, we know that the BSA can be defined as
\be\label{eq:dualBSA}
\BSA(\varrho) = \max \{0, -\min_{W \geq - \id } \tr(W \varrho) \} ,
\ee
where the maximization is over the set of entanglement witnesses that are such that $W \geq - \id$. In turn, the SSIs provide a subset of such witnesses, once they are properly rescaled such that they satisfy $W\geq - \id$ as well. 
Specifically, we can rewrite the three relevant inequalities in \cref{eq:GenSSIsFull} as
\be\label{eq:GenSSIsFullBSA}
\begin{aligned}
\min_{c_x, c_y, c_z \in \mathbb R} \  \aver{ (J_x -c_x \id)^2 + (J_y - c_y \id)^2 + (J_z - c_z \id)^2 } - N/2 & \geq 0 , \\
\min_{c_x\in \mathbb R} \ \  \aver{ (N-1)(J_x -c_x \id)^2- J_y^2 - J_z^2 } + N/2 &\geq 0 , \\
\min_{c_x, c_y \in \mathbb R} \ \aver{ (N-1)[(J_x -c_x \id)^2+(J_y -c_y \id)^2]- J_z^2 } - N(N-2)/4 &\geq 0 ,
\end{aligned}
\ee
and thus, we see that each of them can be written as a minimization over a subset of witnesses parametrized by real numbers $c_x, c_y$ and $c_z$. For example, the top one is given by
\be
\va{J_x}_\varrho + \va{J_y}_\varrho + \va{J_z}_\varrho - N/2 = \min_{c_x, c_y, c_z \in \mathbb R} \aver{W_{c_x,c_y,c_z}} ,
\ee
with
\be
W_{c_x,c_y,c_z} = (J_x -c_x \id)^2 + (J_y - c_y \id)^2 + (J_z - c_z \id)^2 - N/2 ,
\ee
being a parametric set of entanglement witnesses (i.e., $\aver{W_{c_x,c_y,c_z}} \geq 0$ must be satisfied by all separable states). 

Thus, the only missing ingredient is to rescale such witnesses so that they satisfy $W_{c_x,c_y,c_z} \geq - \id$, which means that we have to divide it by the modulus of its minimal (negative) eigenvalue. 
In our case, this is straightforward and we get that the following are all entanglement witnesses that provide valid bounds to the BSA:
\be\label{eq:normWitnBSA}
\begin{aligned}
\tilde W_{c_x,c_y,c_z} &= \frac 2 N \left( (J_x -c_x \id)^2 + (J_y - c_y \id)^2 + (J_z - c_z \id)^2\right) - \id  , \\
\tilde W_{c_x,c_y} &= \frac 2 {N(N-2)} \left( (N-1)[(J_x -c_x \id)^2+(J_y -c_y \id)^2]- J_z^2 \right) - \frac 1 2 \id  , \\
\tilde W_{c_x} &= \frac 4 {N^2} \left( (N-1)(J_x -c_x \id)^2- J_y^2 - J_z^2 \right) + \frac 2 {N} \id ,
\end{aligned}
\ee
and therefore, we can find a lower bound to the BSA by scanning all the expectation values of such witnesses and taking the maximum negative value (with opposite sign).
This in turn corresponds to essentially taking the maximum between the following quantities:
\be\label{eq:GenSSIsFullnormBSA}
\begin{gathered}
\frac 2 N \left[ (\Delta  J_x)^2+(\Delta  J_y)^2+(\Delta  J_z)^2 - N/2 \right] , \\
\frac{4}{N^2} \left[ (N-1)(\Delta  J_k)^2-\aver{J_l^2}-\aver{J_m^2} + N/2 \right] , \\
\frac{2}{N(N-2)} \left[  (N-1)[(\Delta  J_k)^2+(\Delta  J_l)^2] -\aver{J_m^2} - N(N-2)/4 \right] ,
\end{gathered}
\ee
also scanning all possible triples of directions $(k,l,m)$ in principle. 

However, a more automatic and straightforward way to do that is by simply considering the matrix $\mathfrak X$ as defined in \cref{eq:coll_matrices}, namely
\begin{align}
(\mathfrak X_\varrho)_{kl}&= \Gamma_\varrho + \tfrac 1 {2(N-1)} \aver{J_k J_l + J_l J_k}_\varrho - \tfrac{N^2}{4(N-1)} \delta_{kl} ,
\end{align}
and take all and only its positive eigenvalues, which are given by the following expression
\be
(\mathfrak X_\varrho)_{kk}= (\Delta J_k)_\varrho^2 + \tfrac 1 {N-1} \aver{J_k^2}_\varrho - \tfrac{N^2}{4(N-1)} ,
\ee
calculated in some optimal directions $k$ (which correspond to the eigendirections of $\mathfrak X$). 
This way, we get what we defined as our spin-squeezing parameter in \cref{eq:SSIspar}, which we then have to rescale by dividing by (the negative of) its minimal possible value: 
\begin{gather}\label{SSIsNormBSA}
\tilde \xi_{SS}(\varrho):= - \frac 1 {\mathcal N_K} \left[ \Tr(\Gamma_\varrho) - \sum_{k=0}^K  \lambda_k(\mathfrak{X}_\varrho)^{\rm pos} - N/2  \right] ,
\end{gather}
where $\lambda_k(\mathfrak{X}_\varrho)^{\rm pos}$ label the positive eigenvalues of $\mathfrak X$ and $K\in  \{0,1,2\}$ is their number (which can be zero, one or two) and
we have multiplied everything by a rescaling factor that is given by
\be\label{eq:NormNKSSI}
\mathcal{N}_K = \frac N 2  - K\frac{N^2}{4(N-1)} + \frac{N(N+2)}{8(N-1)}\,K(K-1),
\ee
that ensures that the parameter corresponds to an optimal witness such that $W \geq - \id$.

In conclusion, we can find a lower bound to the BSA as
\be
\BSA(\varrho) \geq \tilde \xi_{SS}(\varrho) ,
\ee
where the (negative and rescaled) spin squeezing parameter that gives the bound is given in \cref{SSIsNormBSA} with the normalization as in \cref{eq:NormNKSSI}.

\section{Bounds to the Generalized Robustness}\label{app:GR}

Similar to the BSA, we can also consider the generalized robustness (GR) as a distance-based entanglement measure:
\be
\mathcal{E}_{GR} (\varrho) = \min s \in [0, \infty): \frac{1}{1+s} \varrho + \frac{s}{1 + s} \tau = \sigma,
\ee
where $\sigma$ is a separable state and $\tau \geq 0$ is an arbitrary remainder density matrix. For the GR, the distance function is written as 
\be
D_{GR}(\varrho , \sigma)= \min_{s \in [0,\infty)} \tr((1+s)\sigma - \varrho),
\ee
with $s \tau := (1+s) \sigma - \varrho \geq 0$. The dual problem can hence be written as 
\be
\mathcal{E}_{GR}(\varrho) = \max \{ 0, -\min_{W \in \mathcal{M}_{GR}} \tr(W \varrho) \},
\ee
with $\mathcal{M}_{GR} = \{ W \in \mathcal{W} | \mathbb{1}-W \geq 0 \}$.

Following the same reasoning as described in \cref{app:boundfromSSIs}, we can again take the SSIs as our set of witnesses under consideration and rescale them
in order to be in $\mathcal{M}_{GR}$. In this case we obtain:
\be
\mathcal{E}_{GR} (\varrho) \geq \frac{1}{\mathcal{G}_K} \xi_{SS} (\varrho) \quad \text{with} \quad \mathcal{G}_K = \frac{N^2}{4} - K(K-2) \frac{N^2}{4(N-1)}.
\ee
For finding an upper bound, we can use the exact same algorithm as described in \cref{sec:ansatzintro} and \cref{sec:upper_bound_symm}. In practice once more the upper bound becomes tight when the state is separable (i.e., when a separable decomposition is found), while it remains quite loose when the state is entangled. Once more, this is due to the difficulty of finding the ``closest'' separable state (also in this other metric) when the state is entangled. Note also that in the GR metric the distance scales quite differently with $N$, in particular it is suppressed due to the normalization factor $\mathcal{G}_K$ being of $O(N^2)$.

\begin{figure}
    \centering
    \includegraphics[width=.8\textwidth]{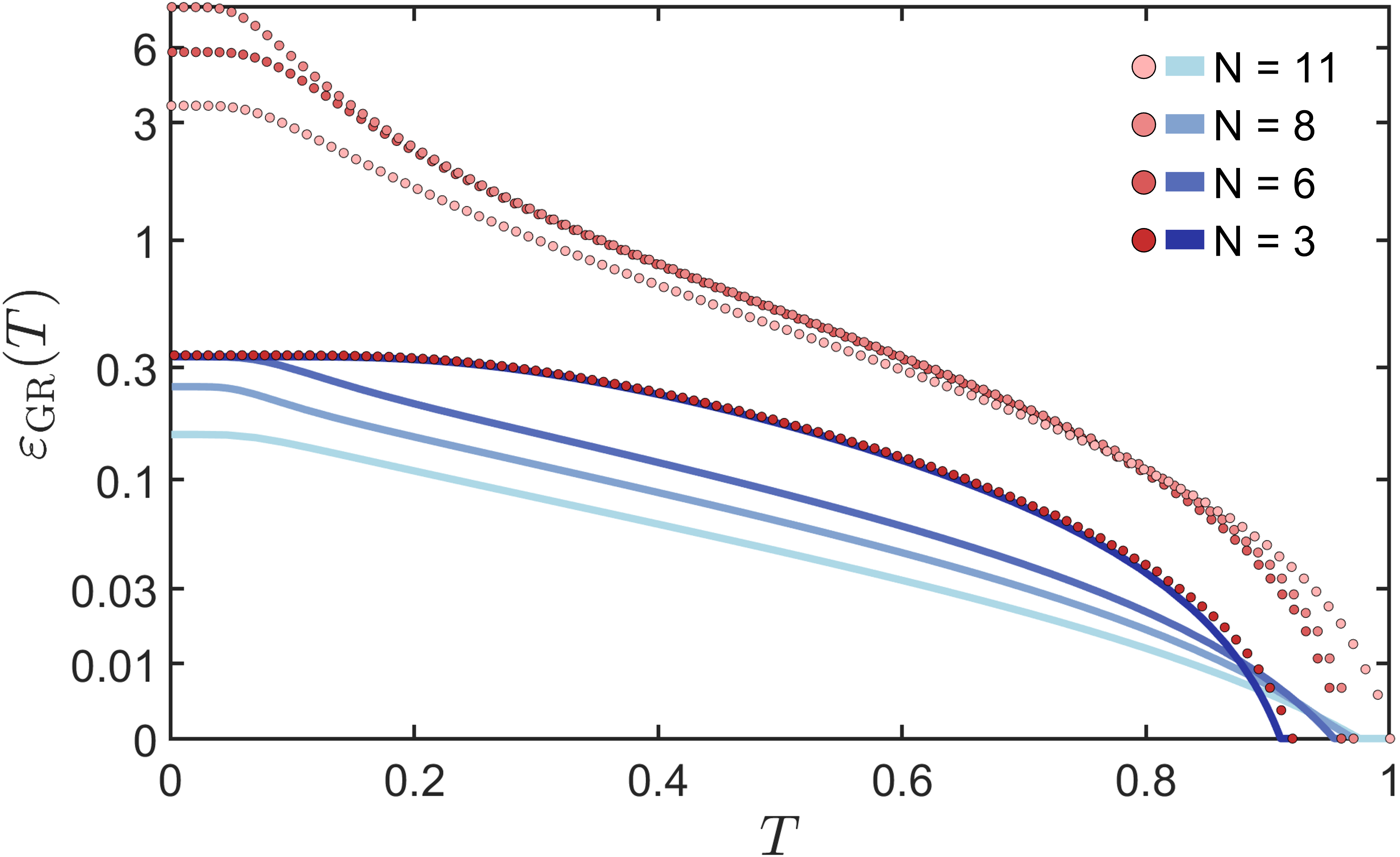}
    \caption{Lower (blue) and upper (red) bounds for the GR of thermal states of the anti-ferromagnetic XXX model.
    The lower bounds depend monotonically on $N$ for $N$ either odd or even but not in general.
    }
    \label{fig:GR_diff_N}
\end{figure}

In \cref{fig:GR_diff_N} we show an exemplary numerical calculation for a thermal state of the antiferromagnetic XXX model.
It is also interesting to point out that for pure states the generalized robustness bounds from above the geometric measure of entanglement~\cite{CavalcantiRobustnessGeometric2006}.

\end{widetext}

\bibliographystyle{quantum}

\bibliography{biblio}

\end{document}